# Grain boundary segregation prediction with a dual-solute model

Zuoyong Zhang and Chuang Deng*

*Department of Mechanical Engineering, University of Manitoba, Winnipeg, Canada MB R3T 5V6*

* Corresponding author: Chuang.Deng@umanitoba.ca

## Graphical abstract

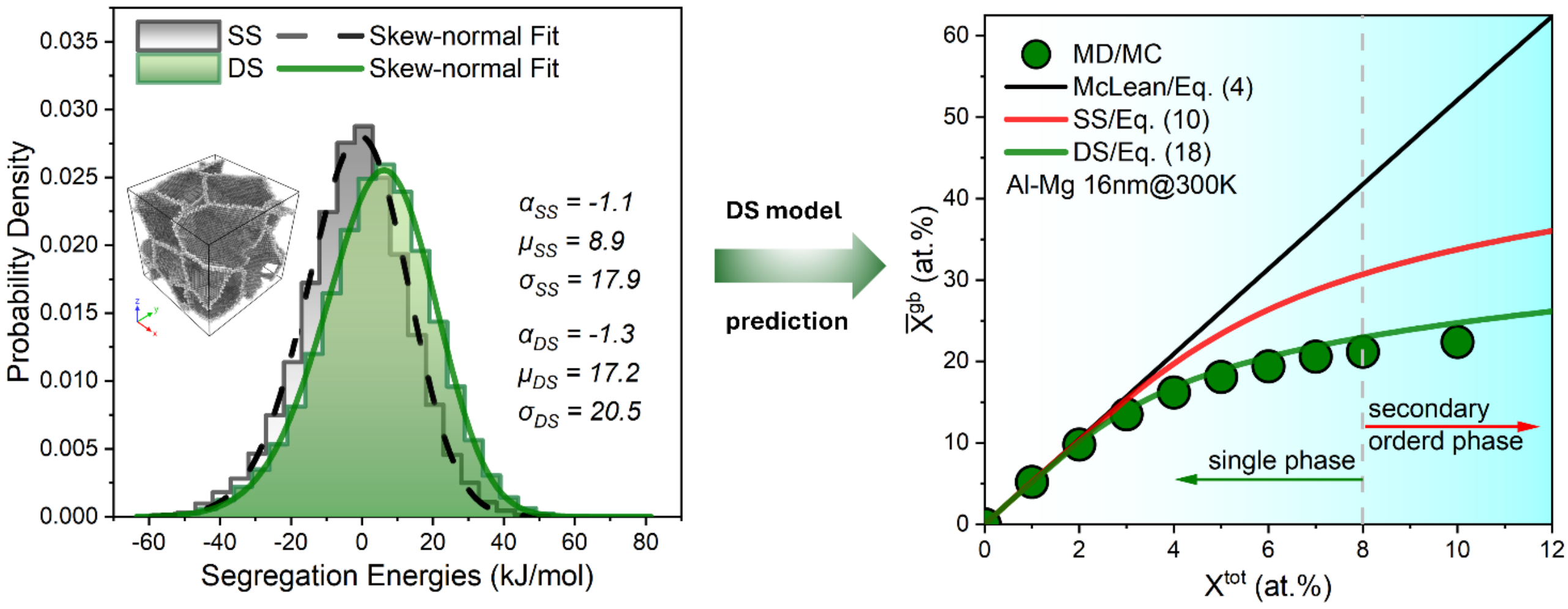

## Abstract

Solute segregation along grain boundaries (GBs) profoundly affects their thermodynamic and kinetic behaviors in polycrystalline materials. Recently, the spectral approach has emerged as a powerful tool to predict GB segregation. However, previous GB segregation predictions using this method relied heavily on single-solute segregation energy spectrum without solute-solute interactions, which were often incorporated through a fitting parameter. In this work, we developed a dual-solute model whose segregation energy spectrum intrinsically incorporates solute-solute interactions. It was first validated for GB segregation prediction in the Al-Mg system, and then extended to several other binary alloy systems. The dual-solute model shows significant improvement over the single-solute model and can accurately predict the real segregation states obtained by hybrid Molecular Dynamics/Monte Carlo simulations within a broad temperature range with different solute concentrations before forming secondary phases. This dual-solute model provides an effective method for accurately predicting GB segregation in nanocrystalline metals.

## 1. Introduction

In polycrystalline metals, fine-grained structures usually exhibit superior mechanical properties compared to their coarse-grained counterparts due to the renowned Hall-Petch strengthening [1–3]. However, these fine structures are often unstable and encounter grain coarsening which can be attributed to the high excess grain boundary (GB) energy that significantly increases the driving force for grain growth in nanocrystalline (NC) metals [4–7]. During the past decades, significant efforts have been made to explore how to stabilize the fine-grained structures and hinder grain coarsening. In this context, solute segregation at GBs has become a promising alloy design tool against grain growth by lowering the excess GB energy and stabilizing the GB network in NC metals [8–14]. Therefore, to effectively manage the process, it is essential to find a reliable method for predicting the solute segregation at GBs.

To predict GB segregation behavior, the GB segregation energy serves as a crucial parameter for influencing the tendency of solute segregation, and that is, negative values indicate solute segregation towards GB regions, while positive values suggest the opposite [12,13]. Initially, the McLean-type approaches [12,15] treated each GB site possessing the same environment with the identical energy state, and the solute segregation driving force was described with a single average segregation enthalpy without considering the solute-solute interactions in it. In the Fowler-Guggenheim [16] (F-G) model, however, a solute-solute interaction component was added to the adsorption energy as a correction for higher solute concentrations. Nevertheless, this separated component was obtained by fitting experimental data [17,18], and still based on the average assumption [19]. Then, several empirical approaches were then developed to describe the segregation energy, such as the Hondros-Seah [20] and Miedema-based ones [21–23]. These classical approaches are very useful for alloy design with plentiful empirical data available to select alloying species [12]. However, the large variations of local atomic environments [24–26] were missing from these approaches due to the average assumption, which may cause significant deviations in GB segregation prediction [27].

A spectral approach, proposed to address the oversimplification of the classical approaches, has garnered significant attention. It can be dated back to 1977 by White and Coghlan [28] who explained the energetic driving force of GB segregation by demonstrating the spectral nature of segregation binding energy. Recently, inspired by the studies of White and Stein [29] and Kirchheim [30,31], Wagih and Schuh [32] developed a spectral approach and proposed a skew-normal model to describe the spectral nature of GB segregation energy. A number of GB segregation energy spectra have been calculated using machine learning techniques by Wagih and Schuh [33,34] providing a basic understanding of the segregation tendency of numerous binary systems.

Since then, there has been significant interest in the spectral approach. Tuchinda and Schuh explored the strong grain size dependencies of solute segregation preference [35] and the triple junction effects on the GB segregation spectra [36]. Then, they calculated the vibrational entropy spectrum for various binary systems and revealed the strong linear correlation between per-site segregation energy and vibrational entropy [37]. By investigating the hydrostatic pressure effects on GB segregation spectra, Zhang and Deng [38] demonstrated that such effects can either enhance or hinder the solute segregation tendencies depending on the alloy systems. Furthermore, they observed a noteworthy transition in segregation tendencies within certain alloy systems, attributable to the changes in the elastic component induced by hydrostatic pressure. Furthermore, Matson and Schuh [39] developed a framework to construct a phase-and-defect diagram based on spectral GB segregation. On the basis of the spectral segregation energy concept, Pal et al. [24] investigated the spectrum of atomic excess free volume in GBs, elucidating the spectral nature of segregation energy from a structure-property correlation aspect.

However, prior GB segregation prediction using the segregation spectrum was only applicable for dilute conditions due to the absence of solute-solute interaction component of segregation energy which may significantly influence the segregation behavior [26,40–42]. To address this issue, two different approaches were developed to incorporate the solute-solute interactions into the skew-normal model: (i) the linear approximation approach which adds a fitting parameter as the solute-solute interaction component [43]; (ii) the atomistic approach which utilizes atomistic simulations to measure the solute-solute interactions by the concept of heat of mixing [44]. The former added a correction to the segregation energy to describe the contribution of solute-solute interactions by fitting hybrid Monte Carlo/Molecular Statics (MC/MS) simulations using the F-G model, which require expensive computing resources. To eliminate the dependence on fitting parameters and reduce the computing resources, Matson and Schuh [44] employed atomistic simulations to measure the solute-solute interactions explicitly for each GB site based on its coordination number and the bonding energies with its nearest neighbors.

These two spectral approaches used hybrid MC/MS simulations as benchmarks for equilibrium state of segregation and assumed the GB network to be static at 0 K. Nevertheless, in realistic materials, GBs may thicken as more solutes segregate [45–47] or undergo significant evolution [48–50] due to the thermal or mechanical effects. These dynamic phenomena of GB network have been well documented in experimental and computational studies but ignored by hybrid MC/MS simulations. Moreover, these two models, relying heavily on the single-solute (SS) segregation energy spectrum, considered the solute-solute interactions not intrinsically but separately from the segregation energies, which may cause underestimation or overestimation of the solute-solute interaction energy. To date, there is still no effective model for predicting GB segregation that accounts for the dynamic evolution of GBs at finite temperatures. Therefore, there is an urgent need for a novel model that can integrate the solute-solute interaction information into the segregation energy spectrum, capturing the dynamic nature of GBs, and enabling accurate prediction of GB segregation at finite temperatures.

In this study, we aim to simplify the segregation prediction model and eliminate the need for empirical parameters and the separate solute-solute interaction term by embedding solute-solute interactions into the segregation spectrum through a dual-solute (DS) segregation framework. The main assumption is that, unless for systems with negligible solute-solute interactions, the energy states that a solute can access would be dramatically changed due to the presence of pre-existing solutes, leading to significant alteration of the shape and position of the segregation energy spectrum. Furthermore, one can also explicitly quantify the solute-solute interactions by comparing the segregation energy spectra based on the SS and DS segregation frameworks. To validate these assumptions, the segregation energy spectra were utilized to forecast GB segregation in the Al-Mg system and extend to other binaries. Compared to other spectral models, the DS model demonstrates superior accuracy in predicting the real GB segregation states as obtained from hybrid MD/MC simulations, which are considerably efficient compared to hybrid MC/MS simulations and capable of capturing the dynamic nature of the GBs during the segregation process.

## 2. Thermodynamics of grain boundary segregation

### *2.1. Interpretation of segregation energy*

It is worth noting that comprehensive thermodynamics treatment [51] is the crucial start point for the discussions of GB segregation isotherms throughout the paper. Hence, the solute segregation energy should be interpreted as the change of free energy, i.e., $\Delta G^{seg}$, between after and before segregation of a solute:

$$\Delta G^{seg} = \Delta H^{seg} - T\Delta S_{vib}^{seg} \tag{1}$$

where T is the absolute temperature of the system and $\Delta S_{vib}^{seg}$ is the excess vibrational entropy of segregation [52]; the segregation enthalpy $\Delta H^{seg}$ is given by:

$$\Delta H^{seg} = \Delta E^{seg} - P\Delta V \tag{2}$$

where $\Delta E^{seg}$ refers to the segregation internal energy; $P\Delta V$ indicates the work done by the pressure P due to a change in volume $\Delta V$. In this work, some necessary simplifications are made as follows: (i) we focus on the segregation phenomenon in solids, where the $P\Delta V$ component is negligible [53], especially when the pressure can be precisely controlled at 0 bar using the Parrinello–Rahman algorithm [54]; (ii) the segregation energies will be calculated at 0 K using the conjugate gradient (CG) algorithm, indicating the negligible contribution of the excess vibrational entropy component. Thus, the segregation free energy can be approximated as [32–34,43,44]:

$$\Delta G^{seg} \approx \Delta E^{seg} \tag{3}$$

### *2.2. Classical segregation isotherms*

One of the most important classical approaches to predict GB segregation is the renowned McLean isotherm [15], where the atomic sites in the polycrystal are divided into two types, e.g., bulk (grain interior) and GB sites, and expressed as:

$$\frac{\bar{X}^{gb}}{1-\bar{X}^{gb}} = \frac{X^{c}}{1-X^{c}} \exp\left(-\frac{\Delta\bar{E}^{seg}}{\kappa_{B}T}\right) \tag{4}$$

where $\bar{X}^{gb}$ is the average solute concentration at GBs, while $X^{c}$ indicates the solute concentration at bulk regions; $\kappa_{B}$ is the Boltzmann constant; and the average segregation energy, $\Delta\bar{E}^{seg}$, is evaluated as:

$$\Delta\bar{E}^{seg} = E^{gb}_{solute} - E^{c}_{solute} \tag{5}$$

where $E^{gb}_{solute}$ and $E^{c}_{solute}$ are the system energies when a solute is sitting at a GB site and bulk site, respectively. For a given system with finite grain sizes and total solute concentrations ($X^{tot}$), both volume fractions for GB ($f^{gb}$) and bulk ($f^{c}$) become finite, where $f^{gb} + f^{c} = 1$, and follow the relationship according to the rule of mixture [55]:

$$X^{tot} = \left(1 - f^{gb}\right)X^{c} + f^{gb}\bar{X}^{gb} \tag{6}$$

Thus, $\bar{X}^{gb}$ can be directly solved for a given $X^{tot}$ in a finite polycrystalline model. However, it consistently fails to predict GB segregation accurately when the solute concentration exceeds the dilute limit due to the oversimplification of the average segregation energy and the absence of solute-solute interactions [27]. The GB volume fraction can be generally described as a function of grain size d and GB thickness t [11]:

$$f^{gb} = 1 - (1 - \frac{t}{d})^{3} \tag{7}$$

GB segregation may show solute concentration dependence when $X^{tot}$ is beyond the dilute limit. To explain this phenomenon, the F-G model [16] added a solute-solute interaction component ($\omega\bar{X}^{gb}$) into the average segregation energy:

$$\frac{\bar{X}^{gb}}{1-\bar{X}^{gb}} = \frac{X^{c}}{1-X^{c}} \exp\left(-\frac{\Delta\bar{E}^{seg}+\omega\bar{X}^{gb}}{\kappa_{B}T}\right) \tag{8}$$

where $\omega$ is the interaction term. Apparently, the solute-solute interaction component increases with the solute concentration at GBs, i.e., $\overline{X}^{gb}$.

As aforementioned, the classical isotherms treat the GB regions as an entity and use a single average energy term to account for evaluation of GB segregation tendency. This simplification excludes the significance of local environment variations, and therefore, may cause significant deviations in predicting GB segregation, especially when the solute concentration is on higher level [27,43].

### *2.3. The spectral approach*

Compared to the classical approaches, the spectral approach suggests that each GB site *i* has its SS segregation energy $\Delta E_i^{SS}$ with a particular probability $F_i^{gb}$ [32], due to the unique local environment:

$$F_i^{gb} = \frac{1}{\sqrt{2\pi}\sigma} \exp\left[-\frac{(\Delta E_i^{SS}-\mu)^2}{2\sigma^2}\right] \mathrm{erfc}\left[-\frac{\alpha(\Delta E_i^{SS}-\mu)}{\sqrt{2}\sigma}\right] \tag{9}$$

where $\alpha$, $\mu$, and $\sigma$ are three fitting parameters referring to the shape (i.e., skewness of the distribution), characteristic energy (i.e., position of the distribution), and width of the skew-normal distribution, respectively. Similar to the form of McLean-type model, the spectral GB segregation prediction model without solute-solute interactions can be expressed as [32]:

$$X^{tot} = \left(1 - f^{gb}\right)X^c + f^{gb} \int_{-\infty}^{\infty} F_i^{gb} \left[1 + \frac{1-X^c}{X^c} \exp\left(\frac{\Delta E_i^{SS}}{\kappa_B T}\right)\right]^{-1} d\Delta E_i^{SS} \tag{10}$$

To incorporate the solute-solute interaction contribution, Wagih and Schuh [43] added an interaction term $\omega\overline{X}^{gb}$ by following the F-G method [16]:

$$X^{tot} = \left(1 - f^{gb}\right)X^c + f^{gb} \int_{-\infty}^{\infty} F_i^{gb} \left[1 + \frac{1-X^c}{X^c} \exp\left(\frac{\Delta E_i^{SS}+\omega\overline{X}^{gb}}{\kappa_B T}\right)\right]^{-1} d\Delta E_i^{SS} \tag{11}$$

where $\omega$ is the solute-solute interaction energy that is obtained by fitting the hybrid MC/MS data using the F-G model. Therefore, it is not derived from the physical properties directly. Then, Matson and Schuh [44] proposed an atomistic method to assess the solute-solute interactions on atomic-level:

$$X^{tot} = \left(1 - f^{gb}\right)X^c + f^{gb}\int_{-\infty}^{\infty} F_i^{gb}\left[1 + \frac{1-X^c}{X^c}\exp\left(\frac{\Delta E_i^{SS} - 2\bar{\Omega}^{gb}\bar{X}^{gb} + 2\Omega^c X^c}{\kappa_B T}\right)\right]^{-1} d\Delta E_i^{SS} \tag{12}$$

where $\bar{\Omega}^{gb}$ and $\Omega^c$ are the average enthalpy of mixing parameters of the GB and bulk regions, respectively, which are derived from atomistic simulations by evaluating the possible bonding energies of a solute atom with its nearest neighbors at the corresponding sites. This approach eliminates the need for fitting parameters but requires additional simulations that exhaustively sample every GB site to obtain $\bar{\Omega}^{gb}$ and every bulk site to calculate the term $\Omega^c$, which is a non-trivial task.

Furthermore, Matson and Schuh [44] proposed a bivariate normal distribution by joining the per-site segregation energy and average mixing enthalpy ($\omega_i$):

$$F_{ij}^{gb} = \frac{1}{\sqrt{(2\pi)^2|\Sigma|}}\exp\left[-\frac{1}{2}(x-\mu)^T\Sigma^{-1}(x-\mu)\right] \tag{13}$$

where $F_{ij}^{gb}$ varies with the vector quantities $x$ and $\mu$; $x$ contains the segregation energy and mixing enthalpy, while $\mu$ contains their means; $\sum$ is their covariance matrix. Based on this bivariate normal distribution, they developed a bivariate model:

$$X^{tot} = \left(1 - f^{gb}\right)X^c + f^{gb}\int_{-\infty}^{\infty}\int_{-\infty}^{\infty} F_{ij}^{gb}\left[1 + \frac{1-X^c}{X^c}\exp\left(\frac{\Delta E_i^{SS} - 2\bar{\Omega}_j^{gb}\bar{X}^{gb} + 2\Omega^c}{\kappa_B T}\right)\right]^{-1} d(\bar{\Omega}_j^{gb})d(\Delta E_i^{SS}) \tag{14}$$

However, this bivariate spectral model significantly increases the computational complexity, and the prediction accuracy was limited at higher solute concentrations with respect to the hybrid MC/MS results.

## 3. Dual-solute segregation framework

### *3.1. Single-solute segregation*

First, we identify all the GB sites of the thermally relaxed NC metal. Next, we construct a neighbor list that includes pairs of potential segregation sites: GB site *i* and their nearest neighbors (site *j*), where site *j* can either be a GB site or a nearby bulk site, indicating that not only GB sites but also near-GB sites can potentially be segregated by solute atoms, which accounts for the dynamic evolution of GBs at finite temperatures. Now, it is ready to calculate the SS segregation energy of each GB site and its nearest neighbors, i.e., $\Delta E_j^{SS}$. This per-site segregation energy is evaluated by:

$$\Delta E_j^{SS} = E_j - E_{solute}^{c} \tag{15}$$

where $E_j$ represents the minimized system energy when a solute is positioned at site *j*, while $E_{solute}^{c}$ denotes the reference energy when a solute is located at a selected bulk site, specifically the center of a large grain to avoid elastic interactions with GBs [32,44]. According to Eq. (15), we can also determine the SS segregation energy ($\Delta E_i^{SS}$) of each GB site *i*, as these are also included in list of site *j*.

### *3.2. Dual-solute segregation*

The DS segregation framework describes the segregation behavior of a couple of solute atoms at or near GBs. It considers solute-solute interactions in a straightforward manner. Now, we need to calculate the segregation energy of GB site *i* ($\Delta E_{i_j}^{seg}$) when a solute atom already exists at one of its nearest neighbors *j*. This can be done by determining the minimized system energy ($E_{ij}$) when a pair of solute atoms exist at the same time and exhausting the whole neighbor list. Then, $\Delta E_{i_j}^{seg}$ can be calculated by:

$$\Delta E_{i_j}^{seg} = \Delta E_{ij}^{seg} - \Delta E_j^{SS} = E_{ij} - E_0 - 2\Delta E^{ref} - \Delta E_j^{SS} \tag{16}$$

Here $\Delta E_{ij}^{seg}$ is the pair-wise segregation energy considering both sites *i* and *j*, which is different from the single-site segregation energy $\Delta E_{i_j}^{seg}$. Eq. (16) indicates that different neighboring solutes result in distinct segregation energies for the same GB site *i*. The term $E_0$ is the system energy of a pure NC metal after relaxation and minimization, while $\Delta E^{ref}$ is a reference energy and evaluated by:

$$\Delta E^{ref} = E_{solute}^{c} - E_0 \tag{17}$$

Up to this point, the solute-solute interactions can be intrinsically incorporated into the DS segregation energy $\Delta E_{i_j}^{seg}$. Then, we notice that $\Delta E_{i_j}^{seg}$ only considers the conditions that another nearest neighboring solute exists but ignores where two solutes are largely separated conditions. This means that its segregation energy spectrum is insufficient for predicting GB segregation. Therefore, we introduce a DS segregation energy spectrum $\Delta E_i^{DS}$, which combines two components: the segregation energy due to solute-solute interactions $\Delta E_{i_j}^{seg}$ and the single-solute segregation energy $\Delta E_i^{SS}$. The first component, $\Delta E_{i_j}^{seg}$, accounts for solute-solute interactions, while the second, $\Delta E_i^{SS}$, represents the segregation energy under dilute conditions, without solute-solute interactions. Here, we do not average the $\Delta E_{i_j}^{seg}$ for GB site *i* over its nearing neighbors, but rather treat them as distinct potential segregation states. We believe that averaging of $\Delta E_{i_j}^{seg}$ would result in the loss of detailed segregation information in the DS spectrum.

In this work, our focus is on the overall GB solute concentration, $\overline{X}^{gb}$, rather than the specific site occupation details. The non-averaged DS segregation energy $\Delta E_{i_j}^{seg}$ and the SS segregation energy $\Delta E_i^{SS}$ indicate that GB site *i* can have up to z+1 energy states, where z is the coordination number of GB site *i*, if site *i* and its neighbors are all occupied by solute atoms. This is because we do not consider scenarios where more than one of its neighboring sites are occupied by solute atoms simultaneously. In this context, we treat the energy state as a probability density, similar to the continuous site occupation concept [56,57]. Consequently, the accumulation of energy states can be used to approximate multiple occupations around site GB *i*.

Then, we employ the DS segregation energy spectrum for GB segregation prediction by following the expression similar to Eq. (10) but intrinsically including the solute-solute interactions in it:

$$X^{tot} = \left(1 - f^{gb}\right)X^c + f^{gb}\int_{-\infty}^{\infty} F_i^{gb}\left[1 + \frac{1-X^c}{X^c}\exp\left(\frac{\Delta E_i^{DS}}{\kappa_B T}\right)\right]^{-1} d\Delta E_i^{DS} \tag{18}$$

Then, Eq. (18) (denoted as the DS model) can be numerically solved for total solute concentration ($X^{tot}$) with respect to bulk solute concentration ($X^c$) after segregation. Finally, the solute concentration at GBs ($\overline{X}^{gb}$) can be derived from Eq. (6).

## 4. Simulation methods

### *4.1. The determination of segregation energies*

The selection of the binary system follows the two criteria: (i) the existence of reliable interatomic potentials; (ii) spectral data availability for comparison in the literature [32,33,43,44]. Accordingly, the Al-Mg system was chosen due to the weak repulsion among solutes [43] and extensive interest in its segregation spectra [32,33,38,43,44]. In this work, we choose the NC structures with randomly generated grains as the study specimens due to their complexity in local atomic environments, which cannot be fully captured by special GBs [58]. The fully relaxed NC Al ($(16\ \mathrm{nm})^3$, Al-16) specimen with randomly oriented grains used for segregation energy calculations, was provided by Wagih and Schuh from their previous studies [32].

In this work, the Large-scale Atomic/Molecular Massively Parallel Simulator (LAMMPS) software package [59] was employed to perform all atomistic simulations, where periodic boundary conditions were assigned in all directions of the simulation box. OVITO [60] was used for visualization and structural analysis. The additive-common neighbor analysis (a-CNA) method [61] was used for structure identification. The non-FCC atoms were all assigned as GB sites. The interatomic interactions in Al-Mg were characterized using the embedded atom method (EAM) [62] potential, which was specifically developed for Al-Mg [63]. This EAM potential has been extensively used in the investigation of segregation in their respective binary systems, yielding reliable results [32,38,43,44]. Then, MS simulations were conducted, and the segregation energies were calculated using Eqs. (15) and (16) for the SS and DS models, respectively.

### *4.2. Hybrid MD/MC simulations*

To evaluate the performance of the DS prediction model, hybrid MD/MC simulations were conducted to obtain the segregated structures at finite temperatures. The simulations were carried out employing a variance-constrained semi-grand-canonical (VC-SGC) ensemble [64], following the procedures described in Refs. [65,66]. MC trails were performed with three constant parameters, i.e., the total solute concentration $c_0$, the chemical potential between the solvent and the solute species $\Delta\mu_0$, and the variance constraint $\kappa$. We used test runs to determine the proper $\Delta\mu_0$ to obtain the desired solute concentration, i.e., $c_0$, with $\kappa = 1000$, which was conducted for all binary systems at 300 K. The optimized parameters are listed in Supplementary Table 1 [67].

The hybrid MD/MC simulations were performed at desired solute concentrations with the consistent timestep of 2 fs. The NC structures were first minimized using the CG algorithm. Then, they were relaxed using the isothermal-isobaric ensemble (NPT) under zero-pressure at 300 K for 200 ps. Thereafter, in each MC cycle, the number of trial moves was set to 10% of the total atoms in the system, with a total of 40,000 MC cycles conducted at 300 K. Additionally, each MC cycle was separated by 10 MD steps at the same temperature to facilitate the structural relaxation after the atomic moves and chemical mixing, where the temperature was controlled using the canonical ensemble (NVT) and the pressure was precisely maintained to 0 bar by a Berendsen barostat [68]. Following this, each NC structure was again relaxed under zero-pressure at 300 K using the NPT ensemble for another 200 ps, then gradually cooled to 0 K with a constant rate of 3 K/ps. Finally, each NC structure was energetically minimized using the CG algorithm with the pressure kept at 0 bar using the Parrinello–Rahman method [54].

### *4.3. Size and temperature effects*

The solute content at GBs may also be affected by the size of the NC structures through tuning the GB volume fraction [11]. Therefore, another large NC-Al ($(40\ \mathrm{nm})^3$, Al-40) was generated using the random Voronoi tessellation with the toolkit Atomsk [69], followed by a thermal relaxation procedure: first, annealing the Al-40 model using the Nose-Hoover thermostat/barostat under 0 bar at 500 K for 500 ps; then, gradually cooling the NC model to 0 K with a constant rate of 3 K/ps, followed by an energy minimization using the CG algorithm. The annealed Al-40 has 18 randomly oriented grains with the average grain size of 18.8 nm and ~3853400 atoms. Subsequently, it was employed for hybrid MD/MC simulations, utilizing the same parameters as those of the Al-Mg ($(16\ \mathrm{nm})^3$) polycrystal at 300 K, to evaluate the size effect on the accuracy of the DS model.

Furthermore, it has been reported that temperature can significantly influence the segregation behaviors of solutes, with higher temperatures leading to lower solute concentrations at GBs [43].

Herein, we performed hybrid MD/MC simulations, as described previously, using the Al-Mg ($(16\ \text{nm})^3$) polycrystal at temperatures of 400 and 500 K. These were carried out to evaluate the potential effects of temperature on the accuracy of the DS prediction model. Test runs were also conducted to find out the optimized $\Delta\mu_0$ for each desired solute concentration and temperature. The resulting parameters are listed in Supplementary Tables 2-3 [67].

## 5. Simulation results

### *5.1. Segregation energy spectra*

The thermally annealed NC Al-16 model has 247,460 atoms and contains 10 randomly oriented grains with an average grain size of 9.1 nm. Its GB network is shown in Fig. 1(a), whose volume fraction is about 18.6%, as determined by the atomic fraction of the non-FCC atoms. Fig. 1(b) presents the SS and DS segregation energy spectra of the Al-Mg system. The former has no solute-solute interactions, while the latter intrinsically contains the solute-solute interaction information in it. The best fitting parameters using the skew-normal distribution (as described by Eq. (9)) for the SS segregation energy spectrum are $\alpha_{\text{SS}} = -1.1$, $\mu_{\text{SS}} = 8.9$ kJ/mol, and $\sigma_{\text{SS}} = 17.9$ kJ/mol. This distribution matches a larger NC-Al with the size of $(36\ \text{nm})^3$ [43] and a smaller one with the size of $(10\ \text{nm})^3$ [44] very well, indicating the reliability of the segregation energy calculation.

The DS spectrum of the Al-Mg system displayed in Fig. 1(b) was plotted using the same bin size with that of the SS spectrum. Its best fitting parameters are $\alpha_{\text{DS}} = -1.3$, $\mu_{\text{DS}} = 17.2$ kJ/mol, and $\sigma_{\text{DS}} = 20.5$ kJ/mol, respectively. Significant changes in the shape and position of the DS spectra can be observed compared to the SS spectrum. It exhibits wider but lower peak probability density than its SS counterpart. This indicates that the segregation energy with solute-solute interactions varies in a larger range and more scattered, while the SS segregation energy seems more concentrated in a narrow range. For example, the DS segregation energy in Al-Mg ranges from -66 to 84 kJ/mol, which is slightly wider than that of SS segregation energy only ranging from -62 to 58 kJ/mol. This can also be interpreted by the larger $\sigma_{\text{DS}}$ values compared to $\sigma_{\text{SS}}$. The reason of this wider range of DS spectrum can be attributed to its data attributions, which contains not only the SS spectral data but also the segregation energies with solute-solute interactions whose data size is much larger than that of the SS ones.

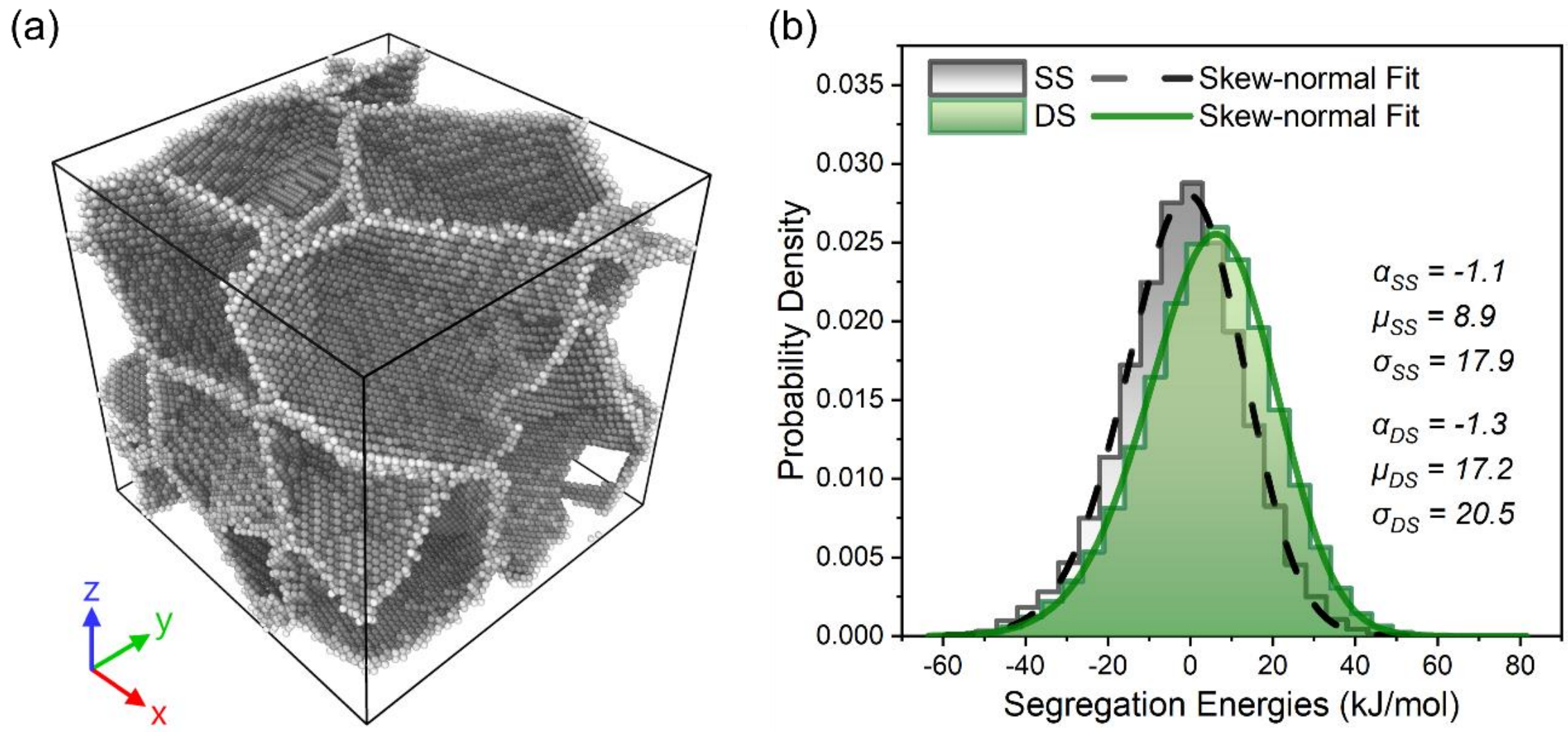

**Fig. 1** (a) The GB network of the thermally relaxed Al-16 NC model with 10 randomly oriented grains of an average grain size of 9.1 nm. (b) The SS and DS segregation energy spectra of Al-Mg with best fitted skew-normal distributions.

The fitting parameter $\alpha$ refers to the shape of the segregation energy spectrum, which indicates the skew direction of the spectrum: positive $\alpha$ means that the spectrum skews to the left, i.e., more negative energies, and vice versa. The $\alpha_{\mathrm{DS}}$ has the same sign but a different value with the corresponding $\alpha_{\mathrm{SS}}$ in the Al-Mg system, suggesting that the solute-solute interactions in the DS model can slightly alter the shape of the spectrum, but not strong enough to change the skew direction of the spectrum. Accordingly, the parameter $\mu$ is the characteristic energy of the spectrum. The greater $\mu_{\mathrm{DS}}$ than $\mu_{\mathrm{SS}}$ in the same system is considered as an indicator of the repulsive solute-solute interactions in the Al-Mg system. It can be perceptually identified by the slight shift to the right side of the DS spectrum compared to the SS spectrum. This means that the segregation energies in the DS spectrum tends to be more positive, suggesting the repulsive interactions among Mg atoms, which has been reported by atomistic studies in the Al-Mg system [43,44].

### *5.2. Segregation states at finite temperatures*

It is necessary to demonstrate the changes in structure and solute distribution of the Al-Mg polycrystal (Al-16) with different solute concentrations after hybrid MD/MC simulations at 300 K, as shown in

Fig. 2. Almost all the solute atoms segregate to GBs, even the Mg content is up to 3 at.%, as shown in Fig. 2(a), indicating the strong segregation tendency of Mg atoms in Al at this condition. Increasing the Mg content to 7 at/%, highly concentrated Mg atoms at GBs can be observed in Fig. 2(b), while many Mg atoms also randomly disperse inside the grain regions. Continuously increasing the Mg concentrations to 8 at.% or higher resulting in the formation of ordered solvent-solute structures, as shown in Fig. 2(c) and (d).

Significant GB network evolution has been captured by the snapshots taken from the same view after hybrid MD/MC simulations, as indicated by the noticeable shifts in the GB network in Fig. 2(b) compared to (a). Additional details can be found in Supplementary Fig. 1 [67]. Higher total solute concentration results in severer GB evolution, likely due to the segregation-induced GB structure transformation [49]. However, the GB network appears static when secondary phases are present within the grain interiors, as shown in Fig. 2(c) and (d), which could be attributed to the pinning effects of the secondary phases on the GB evolution [70,71]. Increasing the Mg concentration from 3 at.% (Fig. 2(a)) to 7 at.% (Fig. 2(b)), the GBs generally become thicker, especially at the triple junctions, which is induced by solute segregation, indicating the possible increase in GB volume fraction, as described by Eq. (7). This phenomenon has also been reported by recent experiments [47]. Therefore, more solute atoms can segregate to the GB regions. At 8 at.% Mg concentration, we observed ordered solvent-solute structures (circled by red ellipse) near GBs, which were identified by the a-CNA method, as shown in Fig. 2(c). After this, higher solute concentrations may not significantly increase the solute contents at GBs, since the formation of these ordered structures can further reduce the system energy [72]. These ordered Al-Mg structures were enhanced by higher Mg contents (e.g., 10 at.%), as shown in Fig. 2(d). Furthermore, these ordered structures are adjacent to GBs, indicating that they start growing from GB edges. The increase in solute atoms at bulk regions suggests that there might be a competition between the segregation of solute atoms at GBs and formation of secondary phases in Al matrix [73]. Similar solute segregation-induced GB evolution and formation of ordered structure in grain interiors can also be observed after hybrid MD/MC simulations of the large Al-40 samples at the same conditions, as shown in Supplementary Fig. 2 [67]. These dynamic evolutions of the GBs at finite temperatures cannot be captured by hybrid MC/MS simulations.

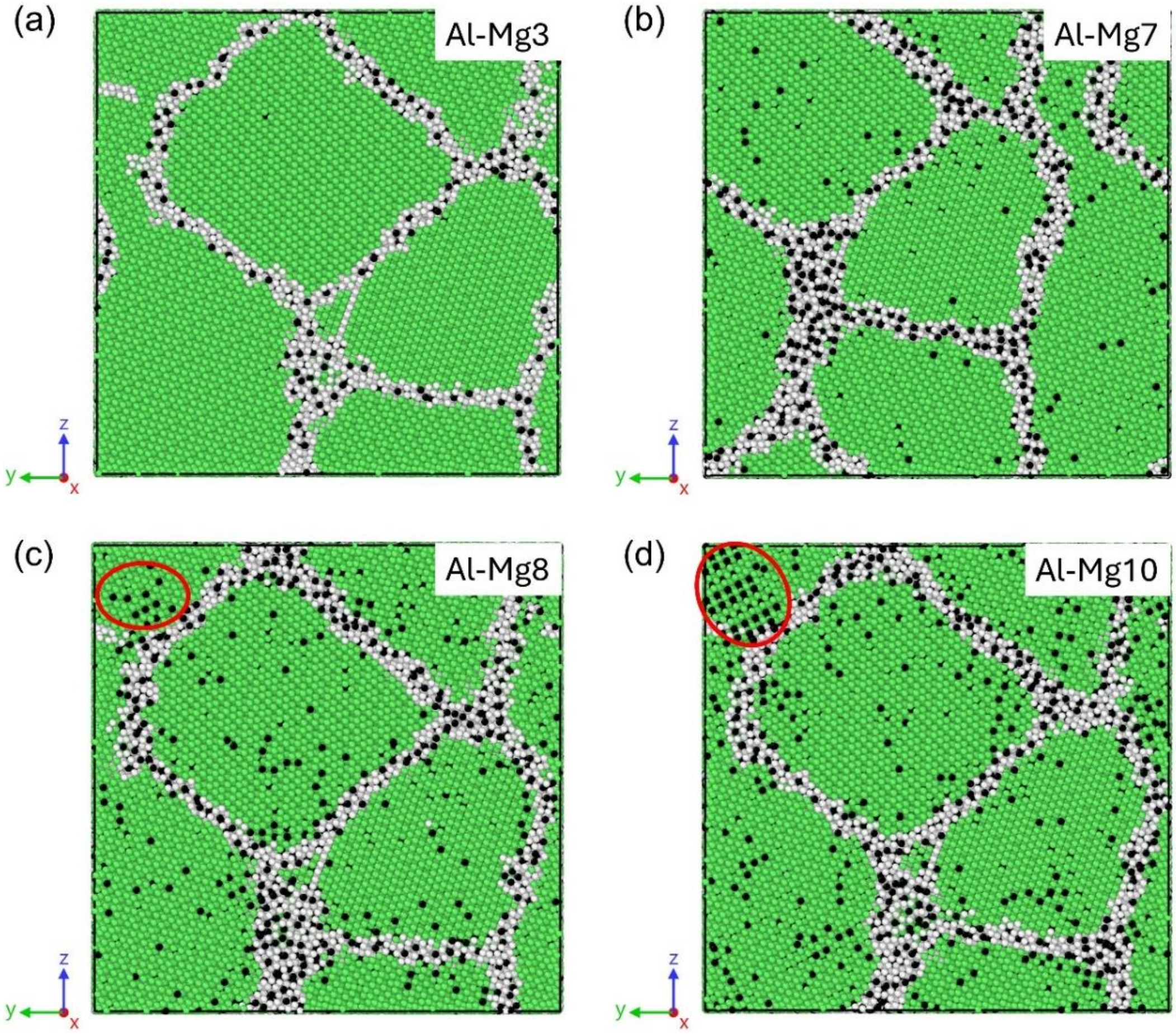

**Fig. 2** The structures and solute distributions of the Al-Mg system with the dimension of $16 \times 16 \times 16\ nm^3$ at different Mg concentrations after hybrid MD/MC simulations at 300 K: (a) Al-Mg 3 at.%, (b) Al-Mg 7 at.%, (c) Al-Mg 8 at.%, and (d) Al-Mg 10 at.%. The green spheres are FCC atoms while the gray ones represent the GB atoms. The dark spheres are the solute atoms. The highlighted regions with red ellipses in (c) and (d) are representatives of ordered solvent-solute structures.

### *5.3. Grain boundary segregation prediction*

The "real" segregation state was obtained by the hybrid MD/MC simulations at finite temperatures using the VC-SGC ensemble [64]. After segregation, each polycrystal was cooled to 0 K. So, these polycrystals can have a common basis for further analysis. The GB solute concentration predicted using the classical McLean model exhibits a linear correlation with the total solute concentration [27,43,44], as shown in Fig. 3(a). It was obtained by solving Eq. (4) with the $\Delta\overline{\mathrm{E}}^{\mathrm{seg}} = -17$ kJ/mol which is the fitting value at 300 K from Ref. [43]. The negative value here is because of the minus sign in front of $\Delta\overline{\mathrm{E}}^{\mathrm{seg}}$ in Eq. (4). No solute-solute interactions were considered in this model. Thus, the prediction of $\overline{\mathrm{X}}^{\mathrm{gb}}$ by the McLean model deviates far away from the hybrid MD/MC results when the total solute concentration is beyond the dilute limit [27,43,44], which can be observed from Fig.

3(a).

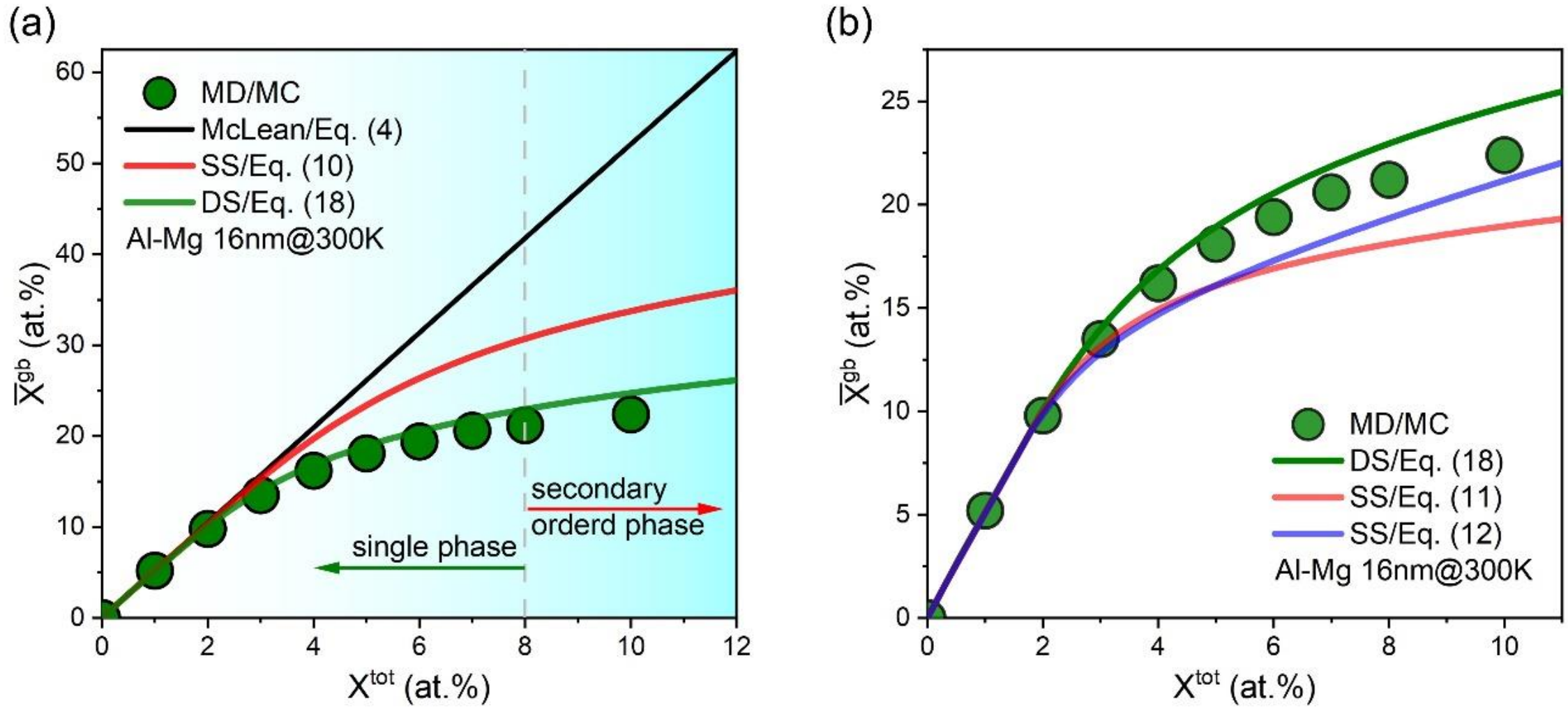

**Fig. 3** (a) GB segregation predictions with respect to the real segregation state (i.e., hybrid MD/MC results ) at 300 K using different models for the Al-Mg system (Al-16): (i) the classical McLean's model Eq. (4), (ii) the spectral SS model Eq. (10) without any solute-solute interactions, and (iii) the spectral DS model Eq. (18) with intrinsically embedded solute-solute interactions. (b) Comparison of predictions between the DS model and different SS models with separated solute-solute interaction terms which are adapted from Refs. [43,44].

Following this, our focus shifts to the SS spectral prediction model without solute-solute interactions. The red solid curve in Fig. 3(a) represents the fitting results by solving the Eq. (10) with the SS spectrum fitting parameters of Al-Mg listed in Fig. 1(b). Compared to the prediction curve of McLean isotherm, it reveals better relative error against the real segregation state, i.e., $\overline{X}^{gb}$ obtained from hybrid MD/MC, indicating the necessary of using the spectral nature of the GB segregation energy. Nevertheless, there is still a dramatic deviation between the curve against the real segregation state, especially at higher Mg concentrations. For instance, the prediction curve largely deviates from the real segregation state when the Mg concentration is greater than 3 at.%. It can be considered as a critical total solute concentration where solute-solute interactions become significant at GBs, with the corresponding $\overline{X}^{gb}$ of 13.5 at.%. This matches well with previous findings that solute-solute interactions are "visible" when solute concentration at GBs is higher than 10 at.% [43]. Obviously, the SS model (Eq. (10)) also fails to predict the real segregation state in Al-Mg at 300 K due to the missing of the critical solute-solute interactions.

Further, the DS prediction (olive solid) curve was obtained by numerically solving Eq. (18) using the spectral parameters of the DS segregation energy spectrum. It effectively predicts the real segregation state at 300 K in the Al-Mg system, as illustrated in Fig. 3(a), even without the separated solute-solute interaction term in Eqs. (11) and (12). This can be attributed to the diverse information of the DS segregation energy spectrum. It intrinsically contains the potential segregation states with and without solute-solute interactions. This allows us to successfully simplify the GB segregation prediction model by eliminating the need for the fitting interaction term $\omega$ in Eq. (11) and the separated interaction terms $\bar{\Omega}^{gb}$ and $\Omega^{c}$ in Eq. (12). Unfortunately, Fig. 3(a) shows that the prediction accuracy of the DS model decreases when the Mg concentration is higher than 8 at.%. The reason for this deviation is probably due to the formation of secondary phases, as shown in Fig. 2(c) and (d). The repulsion between Mg atoms causes solute atoms to preferentially concentrate in bulk regions, which in turn facilitates the formation of ordered solvent-solute structures, thereby reducing the system's energy.

Fig. 3(b) compares the prediction accuracy of the real segregation state at 300 K between the DS model and other SS models that incorporate solute-solute interaction terms. The red solid curve was generated by numerically solving Eq. (11) using the SS spectral parameters in Fig. 1(b), with the interaction term $\omega$ derived from the fitted parameter of the F-G model based on hybrid MC/MS results at 300 K in Ref. [43]. The values of $\bar{\Omega}^{gb}$ and $\Omega^{c}$ values were taken from Ref. [44]. Similarly, using the SS spectral parameters, Eq. (12) was numerically solved, resulting in the blue solid prediction curve shown in Fig. 3(b). Compared to these two SS models, the DS model exhibits superior prediction accuracy within the total solute concentration range before the formation of secondary phases.

### *5.4. Size and temperature effects*

We conducted hybrid MD/MC simulations employing a large NC structure (Al-40) at 300 K to examine the potential effects of the DS model on its prediction accuracy. There are eighteen randomly oriented grains with an average grain size of 18.8 nm, about two times larger than that of the Al-16 sample. The resulting GB structures and solute distributions are shown in Supplementary Fig. 2 [67]. The higher GB solute concentrations can be observed in the large NC model compared to those in the small structure after “real” segregation at 300 K, as shown in Fig. 4(a). This size dependence in $\overline{X}^{gb}$ can be attributed to the relatively smaller GB volume fraction (9.8%) in the larger specimen. As for the prediction accuracy, the DS model can also accurately align with the real $\overline{X}^{gb}$ at 300 K, as shown in Fig. 4(a), indicating the outstanding prediction accuracy for different sizes using the DS model. The notable deviation of the prediction curve of the Al-40 structure from its hybrid MD/MC results in Fig. 4(a) further underscores that ordered solvent-solute structures (Supplementary Fig. 2(d) [67]) have a substantial impact on prediction accuracy, likely due to the competing between secondary phase formation and solute segregation at GBs.

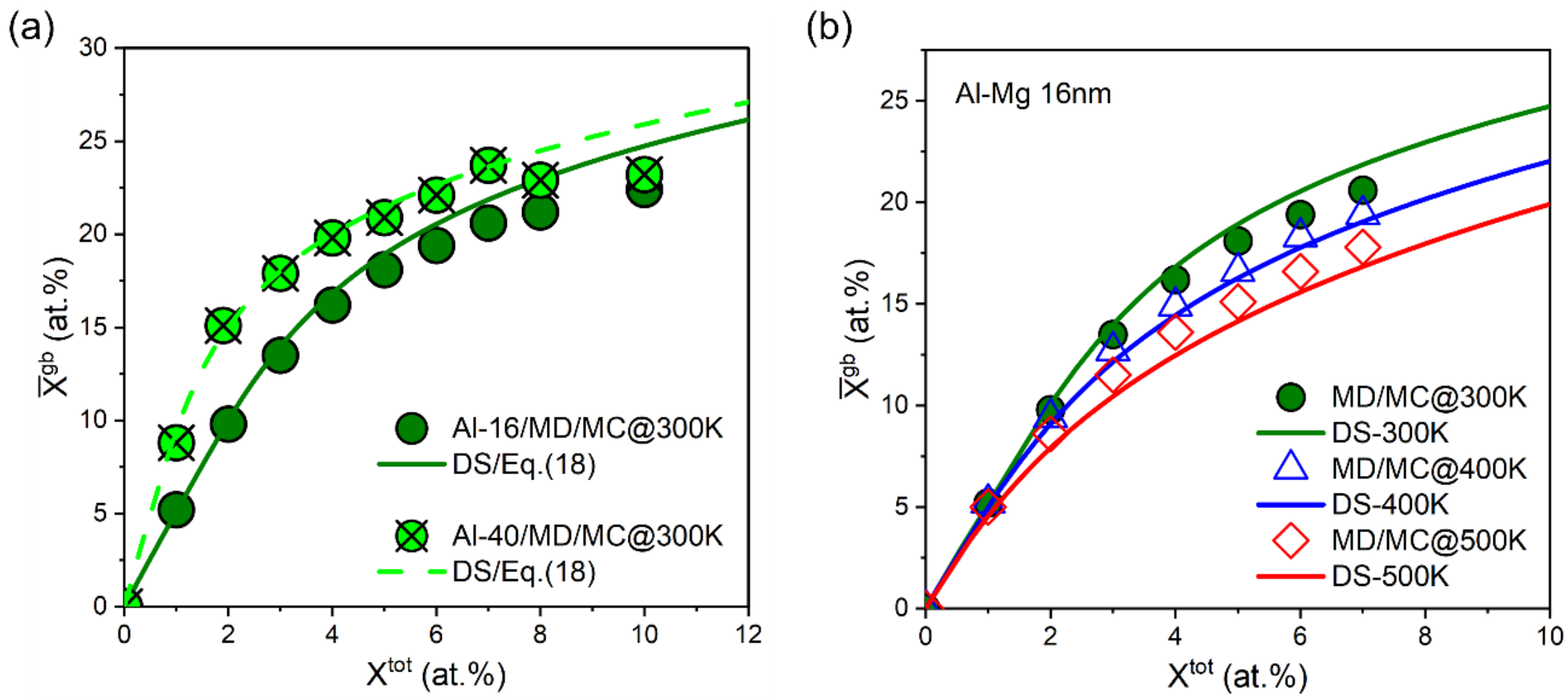

**Fig. 4** (a) GB segregation prediction against hybrid MD/MC results at 300 K using the DS model Eq. (18) in the Al-Mg system for different sizes. (b) The predictions against hybrid MD/MC simulations at three different temperatures (300, 400, and 500 K) in the Al-Mg system.

Grain size dependence of solute segregation has been observed by experiments with larger grain sizes resulting in higher solute concentration at GBs [55,74] due to the smaller GB area over volume ratio. These findings confirm our observation that $\overline{X}^{gb}$ shows a strong size dependence. Nevertheless, the DS model prediction curves also exhibit a size dependence and match well with the real $\overline{X}^{gb}$. Tuchinda and Schuh [35] reported a strong size effect on the McLean segregation energy with a deviation of 6 kJ/mol observed upon increasing the grain size from 5 nm to 40nm. However, we used the same spectral parameters and solute-solute interaction term in both predictions. The prediction accuracy is reliable in both cases. This suggests that the DS model can successfully predict real segregation in those NC structures with different grain sizes just by altering the values of $f^{gb}$ in the DS model in response to the changes in grain size.

Temperature is another crucial factor that may affect solute segregation at GBs. Previous studies reported that the chemical composition at GBs strongly depends on the annealing temperatures, which can be attributed to the differences in segregation enthalpies [75,76] and segregation free energies [77]. This temperature dependence of solute segregation was also confirmed by hybrid MC/MS simulations [43]. To find out if temperature affects the DS prediction accuracy, we conducted hybrid MD/MC simulations at different temperatures in Al-Mg using the Al-16 samples. Indeed, the real $\overline{X}^{gb}$ demonstrates a notable temperature dependence with higher temperatures leading to lower GB solute concentrations at the same total solute concentration, as shown in Fig. 4(b). The temperature dependency becomes evident when the total solute concentration exceeds 1 at.%. At higher total solute concentrations, there is a larger discrepancy in GB solute concentration among different temperatures.

For the 300 and 400 K conditions, the DS model prediction curves exhibit superior accuracy, as shown in Fig. 4(b) and Supplementary Fig. 3 [67]. However, larger prediction errors can be observed for the 500 K conditions. This discrepancy may stem from the artificial nature of the temperature dependence in Eq. (18), as all parameters used here are derived from the 0 K segregation energy spectra, where entropic contributions were ignored [37]. To address this issue, additional atomistic simulations [37,78] may be necessary to account for entropic contributions. Another possible explanation could be the more pronounced GB evolution at elevated temperatures [79], which may be addressed by applying a varying GB volume fraction concept. This approach will be further discussed in the Ag-Ni system in the next section. Nevertheless, the maximum prediction error using the DS mode at 500 K is less than 6%, which is still acceptable in engineering applications. This indicates the reliability of the DS segregation prediction across different temperatures.

**6. Validation in other systems**

After validating the DS model in the Al-Mg system, its application is extended to other systems. Using the same alloy system selection criteria, we have chosen four binary alloy systems: Ag-Cu, Ag-Ni, Al-Ni, and Pt-Au. The interatomic interactions in these binaries are described by the EAM potentials [80–83]. Previous studies have reported solute clustering in Ag-Cu [84] and Pt-Au [83,85], indicating the strong solute-solute attractions in these systems. Therefore, it is necessary to examine the solute clustering effect on the prediction accuracy of the DS model in such binaries.

As discussed in the Al-Mg system, secondary phases can form when the solute concentration exceeds a critical value, which can decrease the prediction accuracy of the DS model. In the Al-Ni system, secondary phases can form even at low Ni concentrations [86], making it a suitable candidate to double-check the precipitation effects on the DS model's prediction accuracy. As for the Ag-Ni system, solute segregation-induced amorphization of interfaces [65] is an interesting phenomenon that may also affect the GB segregation of solute atoms.

To achieve these aims, we constructed the NC-Ag for the Ag-Cu and Ag-Ni systems, and NC-Pt for the Pt-Au system. Then, the NC-Ag and NC-Pt structures were thermally relaxed at 500 K and 900 K using the NPT ensemble for 500 ps, respectively. After thermal relaxation, each NC structure was gradually cooled to 0 K with a constant rate of 3 K/ps. The resulting NC-Ag has a dimension of $16 \times 16 \times 16\ \mathrm{nm}^3$, with 8 randomly oriented grains averaging 9.7 nm in size. The cooled NC-Pt measures $15 \times 15 \times 15\ \mathrm{nm}^3$, with 7 randomly oriented grains of the average size of 9.5 nm. The Al-16 NC structure used for Al-Mg was also utilized for the Al-Ni system here. Subsequently, MS simulations were conducted to calculate the GB segregation energy spectra. Additionally, hybrid MD/MC simulations were employed to determine the real segregation state of these binaries at different solute concentrations at 300 K. The parameters for hybrid MD/MC simulations for each system are listed in Supplementary Tables 4-7 [67].

### *6.1. Strong solute-solute attraction*

Fig. 5(a) and (b) show the structures and solute distributions of Ag-Cu with different solute concentrations after segregation at 300 K. In the Ag-Cu system with 4 at.% Cu, all solute atoms segregate to GBs and form small Cu clusters, confirming the strong attractive interactions among Cu atoms, as shown in Fig. 5(a). When the Cu concentration is increased to 10 at.%, more Cu clusters form, and the smaller clusters grow larger due to the increased segregation of solute atoms to the GB regions, as shown in Fig. 5(b). Slight GB network evolution is also observed after segregation at different solute concentrations. However, this GB evolution is much less pronounced compared to what is observed in the Al-Mg system after segregation.

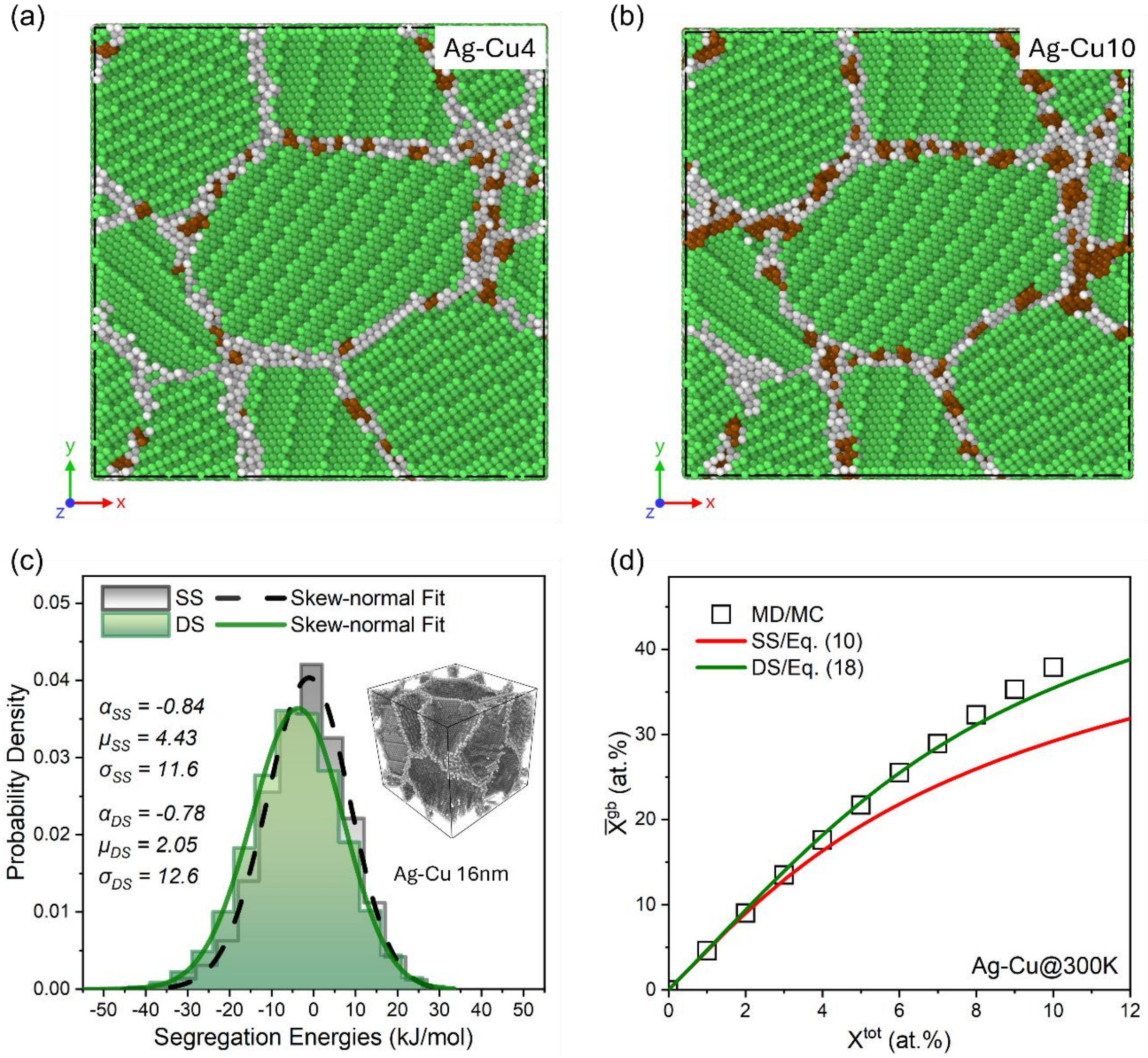

**Fig. 5** GB structures and solute distributions after hybrid MD/MC simulations at 300 K: (a) Ag-Cu 4 at.%, and (b) Ag-Cu 10 at.%. The brown spheres are Cu atoms. (c) the SS and DS segregation energy spectra with corresponding best fitted skew-normal parameters. The inset in (c) shows the relaxed GB network of the NC-Ag with the volume fraction of 20.9%. (d) The prediction results of the SS and DS models against the real segregation state obtained by hybrid MD/MC simulations at 300 K.

Unlike the spectral pattern of Al-Mg, the DS spectrum for the Ag-Cu system slightly shifts to the left side due to more negative segregation energies compared to the SS spectrum, as shown in Fig. 5(c). This behavior of the DS spectrum indicates attractive interactions among Cu atoms in the system. Additionally, Fig. 5(c) shows a broader range of segregation energies in the DS spectrum, resulting in a larger $\sigma$ value compared to the SS spectrum. The prediction results using the SS and DS spectra are shown in Fig. 5(d). These results indicate that the DS model successfully predicts the real segregation state at 300 K. In contrast, the SS model fails due to the absence of solute-solute interactions. Furthermore, the SS prediction curve is lower than the DS curve, confirming the presence of solute-solute attraction in the system.

The GB structures, solute distributions, segregation energy spectra, and prediction results for the Pt-Au system are shown in Supplementary Fig. 4 [67]. Similar solute clustering phenomena occur in the Pt-Au system after hybrid MD/MC simulations at 300 K, as reported by previous studies [83,85]. Compared to the Cu clusters in the Ag-Cu system, the Au clusters in Pt-Au are fewer in number but much larger in size. The strong solute-solute attraction and clustering tendency cause Au atoms to be highly concentrated at GBs. According to the solute distribution analysis after hybrid MD/MC simulations at 300 K, almost all Au atoms segregate to GBs, even with a total solute concentration as high as 8 at.% in the system. Therefore, the hybrid MD/MC data points appear as a linear correlation between the $\overline{X}^{gb}$ and $X^{tot}$. Nevertheless, this phenomenon may not be thoroughly reflected by the spectral pattern, where the DS spectrum only slightly moves to the left side compared to the SS spectrum. This suggests that the DS spectrum may not be able to fully capture the Au-Au attractions at GB regions, which likely explains why the DS prediction curve deviates from the real segregation state when the total solute concentration exceeds 6 at.%. However, the DS prediction curve generally aligns with the real segregation state with acceptable accuracy. Furthermore, the lower SS prediction curve again confirms the strong solute-solute attraction among Au atoms in the system.

### *6.2. Formation of ordered structures*

Previous work [86] has reported that secondary phases can be formed even at low Ni concentrations in the Al-Ni system. In this study, we also observed clear ordered solvent-solute structures near the GBs after hybrid MD/MC simulations at 300 K with a total Ni concentration of 4 at.% in the Al-Ni system, as shown in Fig. 6(a). The navy spheres represent the Ni atoms, while the grey spheres are solvent atoms with FCC atoms removed. Almost all Ni atoms segregate to GB regions with a total solute concentration at 4 at.%, indicating a strong segregation tendency of Ni in Al polycrystals. In the meantime, the solute concentration in bulk regions ($X^{c}_{Ni}$) is only ~0.3 at.%. Similar ordered structures may also be formed prior this Ni concentration after segregation, but it is difficult to identify them using the aCNA method.

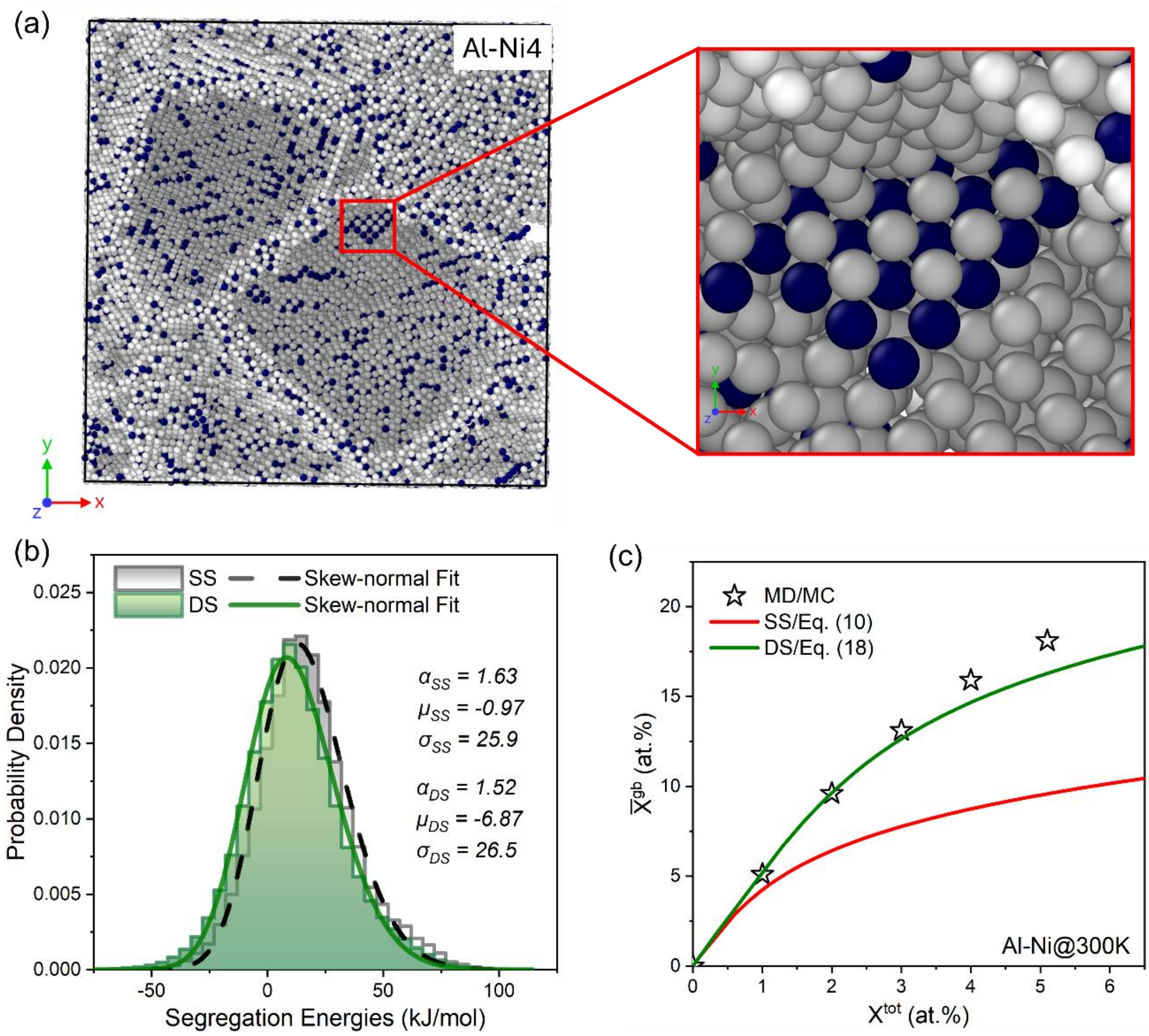

**Fig. 6** (a) Formation of ordered solvent-solute structure near GBs after hybrid MD/MC simulations at 300 K in the Al-Ni with the Ni concentration of 4 at.%. The navy spheres are Ni atoms. (b) The SS and DS segregation energy spectra with best fitted skew-normal distribution parameters for the Al-Ni system. (c) Prediction results against the real segregation states obtained by hybrid MD/MC simulations at 300 K.

Fig. 6(b) shows the SS and DS segregation energy spectra for the Al-Ni system. Slight leftward shift of the DS spectrum compared to the SS spectrum indicates the slight solute-solute attraction. This is further confirmed by characteristic segregation energies of the two spectra: $\mu_{DS} = -6.87$ kJ/mol is smaller than $\mu_{SS} = -0.97$ kJ/mol. The prediction results against the real segregation states are shown in Fig. 6(c). Without solute-solute interactions, the SS model described by Eq. (10) fails to predict the real segregation state. In contrast, the DS model agrees very well with the results of hybrid MD/MC simulations at 300 K, even though the prediction curve slightly deviates from the real $\overline{X}^{gb}$ when the Ni concentration exceeds 4 at.%, which can be explained by the formation of secondary phases. Generally, the DS model performs well in predicting the GB segregation in Al-Ni at finite temperatures.

### *6.3. Segregation-induced amorphization*

Recently, the immiscible Ag-Ni system has garnered extensive research interest due to the strong tendency of Ni segregation at GBs [26,65,66,87]. However, no attempts have been made to predict the GB segregation behavior of Ni in this system. It has been reported that the segregation of Ni in the Ag-Ni system exhibits a fully heterogeneous behavior [65,66]. Ni atoms concentrate at specific GB regions, such as triple junctions, and form large amorphous Ni clusters [65]. This is a distinctive segregation behavior compared to previously discussed binary systems. The partial amorphization at GBs induced by heterogeneous segregation of Ni may profoundly influence the GB volume fraction after solute segregation at finite temperatures.

Fig. 7(a) and (b) illustrate the GB structures and solute distributions after hybrid MD/MC simulations at 300 K in the Ag-Ni system with the total Ni concentration of 1 at.% and 10 at.%, respectively. Nearly all Ni atoms segregate to GB regions in both concentration conditions. However, heterogeneous segregation is much more pronounced in higher total solute concentration conditions, for instance 10 at.%, compared to the real segregation state with only 1 at.% of Ni atoms in the system. This heterogeneous segregation of Ni induces obvious thickening of those GB regions where Ni atoms are concentrated. In other words, this heterogeneous segregation of Ni leads to partial amorphization of the GBs in the Ag-Ni system. Meanwhile, the GB volume fraction significantly increases from 20.9% at a total Ni concentration of 1 at.% to 28.1% when the total solute concentration is 10 at.%. The outstanding changes in GB volume fraction may increase the difficulty in predicting segregation behaviors in the Ag-Ni system. Additional parameters may need to be incorporated to account for the increase in GB volume fraction at higher total solute concentrations.

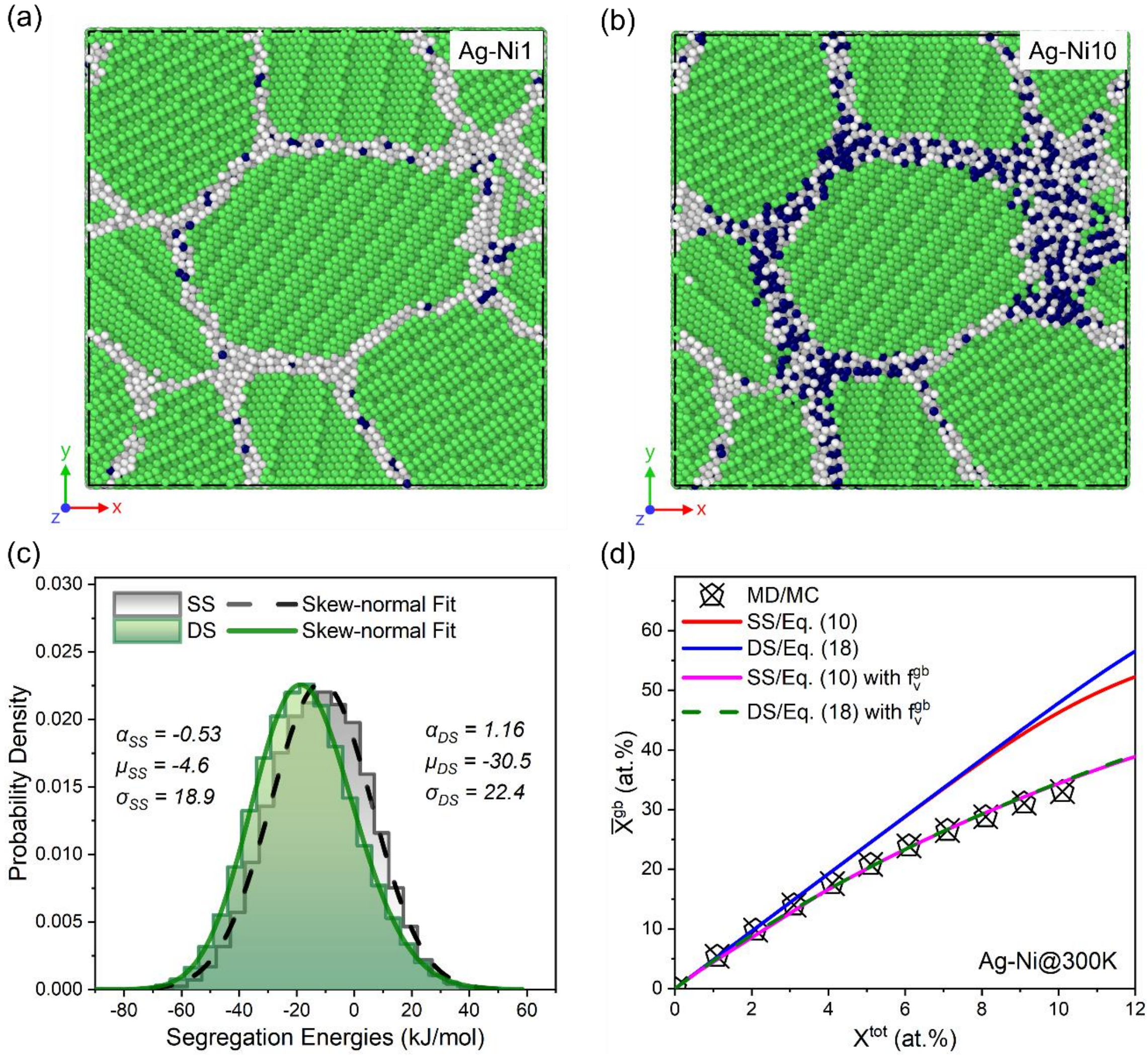

**Fig. 7** GB structures and solute distributions after hybrid MD/MC simulations at 300 K in the Ag-Ni system with the total Ni concentration of (a) 1 at.% and (b) 10 at.%. The navy spheres are Ni atoms. The green spheres represent FCC atoms, while the grey ones are others. (c) The SS and DS spectra for the Ag-Ni system with best fitted skew-normal distribution parameters. (d) Prediction results for the Ag-Ni system against the hybrid MD/MC simulations at 300 K.

Compared to the Ag-Cu and Al-Ni systems, the leftward shift of DS spectrum with respect to the SS spectrum is more significant in the Ag-Ni system, as shown in Fig. 7(c). The solute-solute interactions are even strong enough to alter the shape of the spectrum: the skew direction changes from rightward of the SS spectrum to leftward of the DS spectrum, as indicated by the changes in the sign of $\alpha$ values for the spectra. This change also induces the reduction of characteristic segregation energies from -4.6 kJ/mol to -30.5 kJ/mol. However, the prediction results of the SS (red solid curve) and DS (blue solid curve) models unsurprisingly deviate significantly from the real segregation state, as shown in Fig. 7(d). Now, it is necessary to introduce a concept of varying GB volume fraction ($f_v^{gb}$) to address this issue. We assume that $f_v^{gb}$ is linearly related to the total solute concentration. Then, the expression of $f_v^{gb}$ can be written as:

$$f_v^{gb} = f_0 + k_s X^{tot} \tag{19}$$

where $f_0$ is the GB volume fraction (20.9%) of the pure Ag polycrystal after thermal relaxation at 0 K; $k_s$ is the slope of the linear correlation between $f_v^{gb}$ and $X^{tot}$, whose value is 0.82, as shown in Supplementary Fig. 5 [67]. With correction of $f_v^{gb}$, both SS and DS models align very well with the real segregation states of $\overline{X}^{gb}$. The two prediction curves even coincide with each other, as shown in Fig. 7(d). This indicates that the varying GB volume fraction used in predicting GB segregation is more important than the prediction model themselves in the Ag-Ni system, due to the solute segregation-induced amorphization at GB regions.

Here, we notice that $k_s$ is an empirical parameter. To simplify Eq. (19), we examined and showed the correlation between $f^{gb}$ and $X^{tot}$ for larger NC-Ag model, as shown in Supplementary Fig. 6(a) [67]. The results showed that GB volume fraction quasi-linearly increases with $X^{tot}$ in different temperatures. The slope $k_s$ showed strong temperature dependence. It increases from 0.9 at 300 K to 1.37 at 700 K. However, we found that if we approximate the slope $k_s = 1$ as a constant, the DS model can also accurately predict the segregation behaviors in the Ag-Ni system at different temperatures, as shown in Supplementary Fig. 6(b) [67]. Therefore, the slope $k_s$ can then be treated as a constant, eliminating the needs for empirical parameter $k_s$. The Eq. (19) can then be rewritten as:

$$f_v^{gb} = f_0 + X^{tot} \tag{20}$$

## 7. Discussions

### *7.1. Comparison with other spectral approaches*

Previously, we introduced two other spectral models for GB segregation prediction that include separate solute-solute interactions to account for conditions where the total solute concentration exceeds the dilute limit [43,44]. Both approaches use the SS segregation energy spectrum for segregation prediction, where solute-solute interactions are absent. Hybrid MC/MS simulations were considered the true segregation state, which treated GBs as static at 0 K. Therefore, the GB evolution in Al-Mg or the partial GB amorphization induced by solute segregation at finite temperatures reported in this work could not be captured by hybrid MC/MS simulations. This GB evolution and amorphization may originate from two reasons: (i) solute atoms segregating to near GB sites after exhausting all available segregation GB sites, and (ii) segregation-induced GB phase (or structural) transformations [49]. Further investigations will be needed to isolate the origin of the GB evolution and amorphization in the future.

Moreover, the linear approximation approach relies on fitting the hybrid MC/MS results to account for solute-solute interactions, rather than physically deriving them [43]. This hybrid MC/MS method is extremely computationally expensive as described in Ref. [44]. The atomistic approach computed the solute-solute interactions by approximating the heat of mixing between different bond types. Both studies have shown considerable prediction accuracy against the hybrid MC/MS results. Although the atomistic approach can quickly determine those parameters by sampling part of the GB sites, the DS model performs better in predicting realistic segregation behaviors in binary systems at finite temperatures with outstanding prediction accuracy.

It is worth noting that, unlike the atomistic approach described by Eq. (12) [44], the DS model incorporates solute-solute interactions at GBs but does not embed the solute-solute interactions in bulk regions (i.e., the term $\Omega^c X^c$). We believe that there is also a threshold of solute concentration before solute-solute interactions in bulk regions become significant, as discussed for the solute-solute interactions in GBs [43]. These interactions seem negligible for GB segregation prediction before the formation of secondary phases in bulk regions, such as when the Mg concentration is equal or lower than 7 at.%. That is likely why Eq. (12) overestimates the solute-solute interactions before the formation of secondary phases but matches better with hybrid MD/MC simulations after forming secondary phases, as shown in Fig. 3(b). Nevertheless, the bulk solute-solute interactions after the formation of secondary phases are out of the scope of this work and need further investigations in the future.

*7.2. The solute-solute interactions in the DS model*

The solute-solute interactions, specifically at GBs, are intrinsically embedded in the DS model by integrating the DS segregation energy spectrum. For further comparison with the atomistic approach by Matson and Schuh [44], it is necessary to isolate the solute-solute interactions from the SS and DS segregation spectra for the Al-Mg system using the following expression:

$$\Delta E_{ij}^{\mathrm{int}} = \Delta E_{i}^{\mathrm{SS}} + \Delta E_{j}^{\mathrm{SS}} - \Delta E_{ij}^{\mathrm{seg}} \tag{21}$$

where $\Delta E_{ij}^{\mathrm{int}}$ is the solute-solute interaction energy between GB site *i* and one of its nearest neighbors, site *j* derived from the three types of segregation energies. We do not average the $\Delta E_{ij}^{\mathrm{int}}$ for GB site *i* but rather focus directly on its distribution. We believe that averaging may cause the loss of detailed information in solute-solute interaction energy, and further induce deviations in segregation prediction. Each bonding type has its unique segregation probabilities. This interaction energy is physically similar to the calculated heat of mixing in Ref. [44] but obtained in different manners.

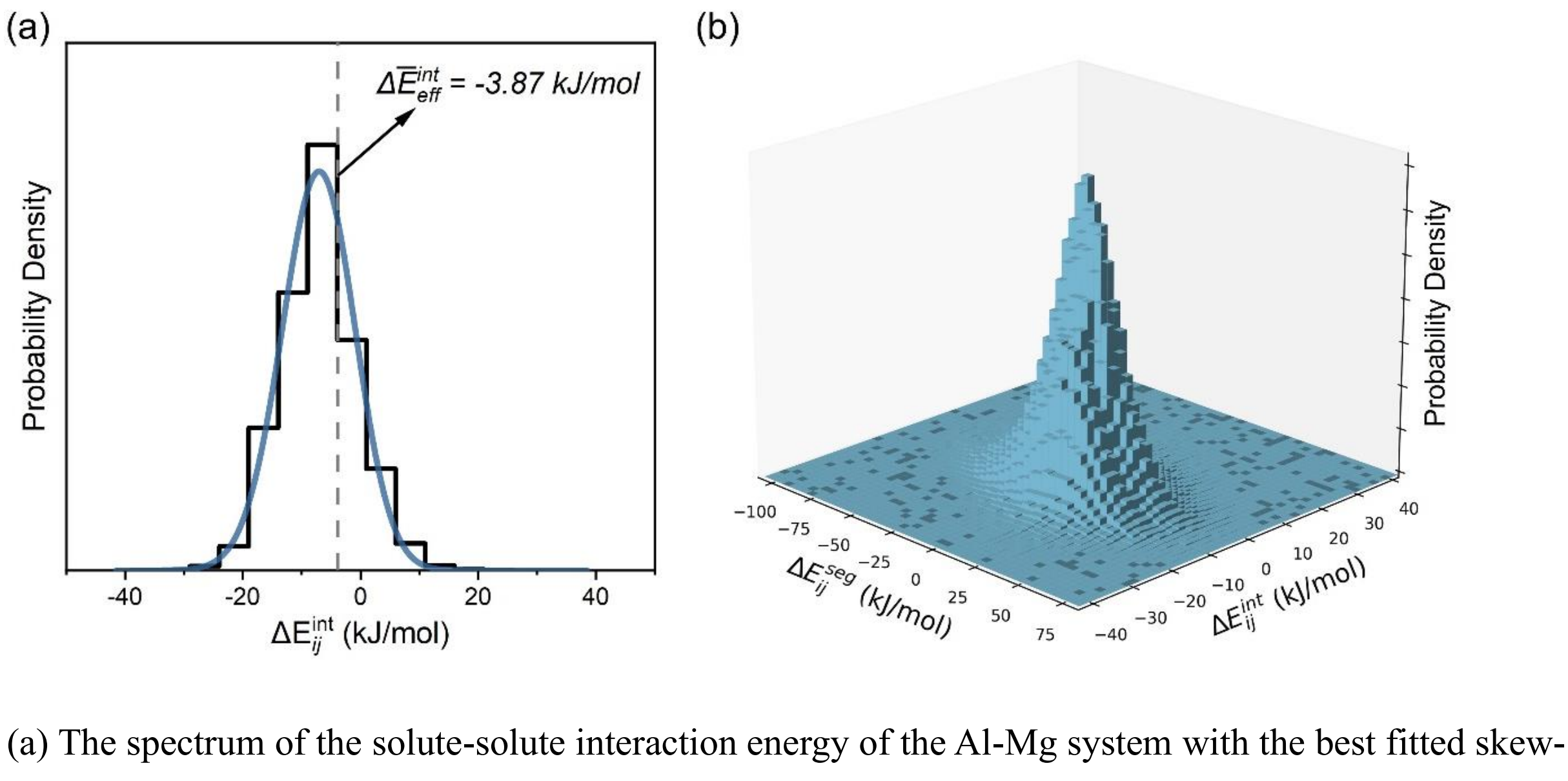

**Fig. 8** (a) The spectrum of the solute-solute interaction energy of the Al-Mg system with the best fitted skew-normal (Eq. (9)) distribution. (b) The bivariate distribution of the pair-wise segregation energy and interaction energy of the Al-Mg system.

The spectrum of the solute-solute interaction energy of the Al-Mg system is shown in Fig. 8(a). It also exhibits a skew-normal type distribution. The colored curve in Fig. 8(a) shows the best fitted skew-normal distribution with a characteristic parameter of -3.87 kJ/mol, which we treat as the effective solute-solute interaction energy for the Al-Mg system at GBs. The value matches well with that of the average per-site parameter of heat of mixing in Ref. [44]. Therefore, it is reasonable to use the interaction parameters from it for comparison in Fig. 3(b). The bivariate skew-normal distribution of the pair-wise segregation energy $\Delta E_{ij}^{seg}$ and interaction energy $\Delta E_{ij}^{int}$ appears more concentrated at the corresponding characteristic positions compared to that of the averaged parameters in Ref. [44]. These positions are crucial for predicting the segregation behavior of the binary system.

### *7.3. Limitations of the DS model and future work*

Although we have shown that the DS model performs very well in predicting GB segregation behaviors in several binary systems, one may argue that it does not include the entropic effect, leading to an artificial temperature dependence in the model. Indeed, the spectral parameters were obtained from MS simulations at 0 K, where vibrational entropy was excluded. However, the maximum prediction error for different temperatures is no more than 6%, which is acceptable for engineering applications. Despite this, there are two reliable approaches available to overcome this entropy-induced prediction error: (i) measuring the vibrational entropy through atomistic simulations [37], and (ii) calculating the segregation free energy directly [78]. Nevertheless, more efforts are needed to incorporate the solute-solute interactions, as such interactions were not defined in either approach.

The DS model is limited to predicting solute segregation at GBs before the formation of secondary phases. During annealing at finite temperatures, competition between forming secondary phases and segregating to GBs may exist. These secondary phases can be formed in bulk, at GBs, or both, which significantly increases the difficulty of GB segregation prediction. The exclusion of solute-solute interactions in bulk regions makes the DS model unsuitable for GB segregation prediction when the solute does not tend to segregate. Additionally, similar to the atomistic approach by Matson and Schuh [44], the DS model only considers the solute-solute interactions of nearest neighbors, while longer range interactions beyond nearest neighbors are missing in this model. Therefore, there may be some discrepancies between realistic state and segregation prediction by the DS model in systems with strong solute clustering tendency, such as the Pt-Au system. As discussed previously, further investigations are needed to perfect the DS model. The bulk and GB solute-solute interactions after the formation of secondary phases, the competition between precipitation and GB segregation, and the segregation in multiple element systems are areas worth further exploration in the future.

## 8. Conclusion

We introduced a DS segregation framework which was used to calculate the segregation energy spectra that inherently considers the solute-solute interactions. Based on this, we developed a DS model for GB segregation prediction. We then showed that the DS model accurately predicts the real GB segregation states obtained by hybrid MD/MC simulations at finite temperatures in several distinct binary alloy systems, e.g., Al-Mg, Ag-Cu, Pt-Au, Al-Ni, and Ag-Ni. The main conclusions of this work can be drawn as follows:

(1) The DS segregation energy spectrum differs from the SS condition in each binary alloy system. It can either shift rightward due to a higher density of positive segregation energies (solute-solute repulsion) or leftward indicating a higher density of negative segregation energies (solute-solute attraction) in the DS spectrum. The skew-normal fitted characteristic energy of the spectra can also serve as the indicator of repulsive or attractive solute-solute interactions. A larger $\mu_{\mathrm{DS}}$ compared to $\mu_{\mathrm{SS}}$ signifies solute-solute repulsion, as shown in Al-Mg, while a smaller $\mu_{\mathrm{DS}}$ than $\mu_{\mathrm{SS}}$ indicates the attractive solute-solute interactions, e.g., Ag-Cu, Pt-Au, Al-Ni, and Ag-Ni.

(2) Hybrid MD/MC simulations accurately represent realistic segregation states at finite temperatures. At higher total solute concentrations, ordered structures were observed in the Al-Mg and Al-Ni systems, which is consistent with previous studies. In the Ag-Cu and Pt-Au systems, many solute-solute clusters formed at GBs after segregation at finite temperatures, reflecting the strong solute-solute attraction, which aligns with the behaviors observed in the DS spectra. In the Ag-Ni system, partial GB amorphization induced by solute segregation was also identified as previous reports. These specific structural features observed after segregation at finite temperatures confirm the reliability of the hybrid MD/MC simulations.

(3) The DS model successfully predicts real segregation states at finite temperatures in different systems with considerable accuracy before the formation of secondary phases. Compared to previous spectral models, the DS model demonstrates superior prediction accuracy in the Al-Mg system. The effects of size and temperature on the DS model's prediction accuracy are acceptable within the scope of engineering applications.

In summary, the DS model is reliable for predicting GB segregation behaviors in various binary systems. Nevertheless, future work is still needed to address bulk and GB solute-solute interactions after the formation of secondary phases, the competition between precipitation and GB segregation, and the segregation in multiple element systems. Despite these areas requiring further exploration, our study presents a novel and simplified method for predicting solute segregation at GBs at finite temperatures with significant accuracy.

**Acknowledgement**

This research was supported by the NSERC Discovery Grant (RGPIN-2019-05834), Canada, and the use of computing resources provided by the Digital Research Alliance of Canada. During the preparation of this manuscript, the authors used ChatGPT to improve its readability. After using this tool, the authors reviewed and edited the manuscript as needed and take full responsibility for the content of the publication.

**Declaration of competing interest**

The authors declare that they have no known competing financial interests or personal relationships that could influence the work reported in this paper.

**Supplementary materials for**

# Grain boundary segregation prediction with a dual-solute model

Zuoyong Zhang and Chuang Deng*

*Department of Mechanical Engineering, University of Manitoba, Winnipeg, Canada MB R3T 5V6*

* Corresponding author: Chuang.Deng@umanitoba.ca

**Supplementary Table 1** Hybrid Molecular Dynamics/Monte Carlo (MD/MC) parameters for the Al-Mg system with $\kappa$ = 1000 were obtained by test runs at 300 K. $c_0$ indicates the desired solute concentration, while $\Delta\mu_0$ is the chemical potential difference between the solvent and solute species. The $\mathrm{X}^{\mathrm{tot}}$ and $\overline{\mathrm{X}}^{\mathrm{gb}}$ are the total solute concentration and grain boundary (GB) solute concentration after hybrid MD/MC simulations, respectively. The colored region in the table below indicates the total solute concentrations that secondary phases can be observed after the hybrid MD/MC simulations.

| $c_0$×100, at.% | 1 | 2 | 3 | 4 | 5 | 6 | 7 | 8 | 10 |
|---|---|---|---|---|---|---|---|---|---|
| $\Delta\mu_0$ | -1.683 | -1.759 | -1.847 | -1.811 | -1.860 | -1.877 | -1.885 | -1.891 | -1.890 |
| $\mathrm{X}^{\mathrm{tot}}$, at.% | 1 | 2 | 3 | 4 | 5 | 6 | 7 | 8 | 10 |
| $\overline{\mathrm{X}}^{\mathrm{gb}}$, at.% | 5.2 | 9.8 | 13.5 | 16.2 | 18.1 | 19.4 | 20.6 | 21.2 | 22.4 |

**Supplementary Table 2** Hybrid MD/MC parameters for the Al-Mg system with $\kappa$ = 1000 were obtained by test runs at 400 K. The $\mathrm{X}^{\mathrm{tot}}$ and $\overline{\mathrm{X}}^{\mathrm{gb}}$ are the total solute concentration and GB solute concentration after hybrid MD/MC simulations, respectively.

| $c_0$×100, at.% | 1 | 2 | 3 | 4 | 5 | 6 | 7 |
|---|---|---|---|---|---|---|---|
| $\Delta\mu_0$ | -1.703 | -1.756 | -1.795 | -1.825 | -1.837 | -1.851 | -1.859 |
| $\mathrm{X}^{\mathrm{tot}}$, at.% | 1 | 2 | 3 | 4 | 5 | 6 | 7 |
| $\overline{\mathrm{X}}^{\mathrm{gb}}$, at.% | 5.2 | 9.4 | 12.7 | 14.9 | 16.6 | 18.3 | 19.4 |

**Supplementary Table 3** Hybrid MD/MC parameters for the Al-Mg system with $\kappa$ = 1000 were obtained by test runs at 500 K. The $X^{tot}$ and $\overline{X}^{gb}$ are the total solute concentration and GB solute concentration after hybrid MD/MC simulations, respectively.

| $c_0$×100, at.% | 1 | 2 | 3 | 4 | 5 | 6 | 7 |
|---|---|---|---|---|---|---|---|
| $\Delta\mu_0$ | -1.715 | -1.749 | -1.775 | -1.792 | -1.801 | -1.827 | -1.837 |
| $X^{tot}$, at.% | 1 | 2 | 3 | 4 | 5 | 6 | 7 |
| $\overline{X}^{gb}$, at.% | 5.0 | 8.6 | 11.5 | 13.6 | 15.1 | 16.6 | 17.8 |

**Supplementary Table 4** Hybrid MD/MC parameters for the Ag-Cu system with $\kappa$ = 1000 were obtained by test runs at 300 K. The $X^{tot}$ and $\overline{X}^{gb}$ are the total solute concentration and GB solute concentration after hybrid MD/MC simulations, respectively.

| $c_0$×100, at.% | 1 | 2 | 3 | 4 | 5 | 6 | 7 | 8 | 9 | 10 |
|---|---|---|---|---|---|---|---|---|---|---|
| $\Delta\mu_0$ | 0.726 | 0.711 | 0.702 | 0.693 | 0.679 | 0.668 | 0.661 | 0.652 | 0.639 | 0.631 |
| $X^{tot}$, at.% | 1 | 2 | 3 | 4 | 4.9 | 6 | 7 | 8 | 9 | 10 |
| $\overline{X}^{gb}$, at.% | 4.5 | 8.8 | 13.1 | 17.3 | 20.4 | 25.1 | 28.5 | 31.8 | 34.8 | 37.6 |

**Supplementary Table 5** Hybrid MD/MC parameters for the Pt-Au system with $\kappa$ = 1000 were obtained by test runs at 300 K. The $X^{tot}$ and $\overline{X}^{gb}$ are the total solute concentration and GB solute concentration after hybrid MD/MC simulations, respectively.

| $c_0$×100, at.% | 1 | 2 | 3 | 4 | 5 | 6 | 7 | 8 |
|---|---|---|---|---|---|---|---|---|
| $\Delta\mu_0$ | -2.415 | -2.402 | -2.396 | -2.389 | -2.381 | -2.365 | -2.343 | -2.312 |
| $X^{tot}$, at.% | 1 | 2 | 3 | 4 | 4.9 | 6 | 7 | 7.9 |
| $\overline{X}^{gb}$, at.% | 5.3 | 10.4 | 15.6 | 21 | 26.3 | 31.4 | 36.7 | 41.2 |

**Supplementary Table 6** Hybrid MD/MC parameters for the Al-Ni system with $\kappa = 1000$ were obtained by test runs at 300 K. The $X^{tot}$ and $\overline{X}^{gb}$ are the total solute concentration and GB solute concentration after hybrid MD/MC simulations, respectively.

| $c_0\times100$, at.% | 1 | 2 | 3 | 4 | 5 |
|---|---|---|---|---|---|
| $\Delta\mu_0$ | 2.261 | 2.189 | 2.156 | 2.081 | 2.053 |
| $X^{tot}$, at.% | 1 | 2 | 3 | 4 | 5.1 |
| $\overline{X}^{gb}$, at.% | 5.1 | 9.6 | 13.1 | 15.9 | 18.1 |

**Supplementary Table 7** Hybrid MD/MC parameters for the Ag-Ni system with $\kappa = 1000$ were obtained by test runs at 300 K. The $X^{tot}$ and $\overline{X}^{gb}$ are the total solute concentration and GB solute concentration after hybrid MD/MC simulations, respectively.

| $c_0\times100$, at.% | 1 | 2 | 3 | 4 | 5 | 6 | 7 | 8 | 9 | 10 |
|---|---|---|---|---|---|---|---|---|---|---|
| $\Delta\mu_0$ | 1.078 | 1.083 | 1.089 | 1.093 | 1.106 | 1.109 | 1.113 | 1.115 | 1.119 | 1.121 |
| $X^{tot}$, at.% | 1.1 | 2.1 | 3.1 | 4.1 | 5.1 | 6.1 | 7.1 | 8.1 | 9.1 | 10.1 |
| $\overline{X}^{gb}$, at.% | 5.1 | 9.4 | 13.4 | 17 | 20.3 | 23.3 | 25.9 | 28.1 | 30.2 | 32.1 |

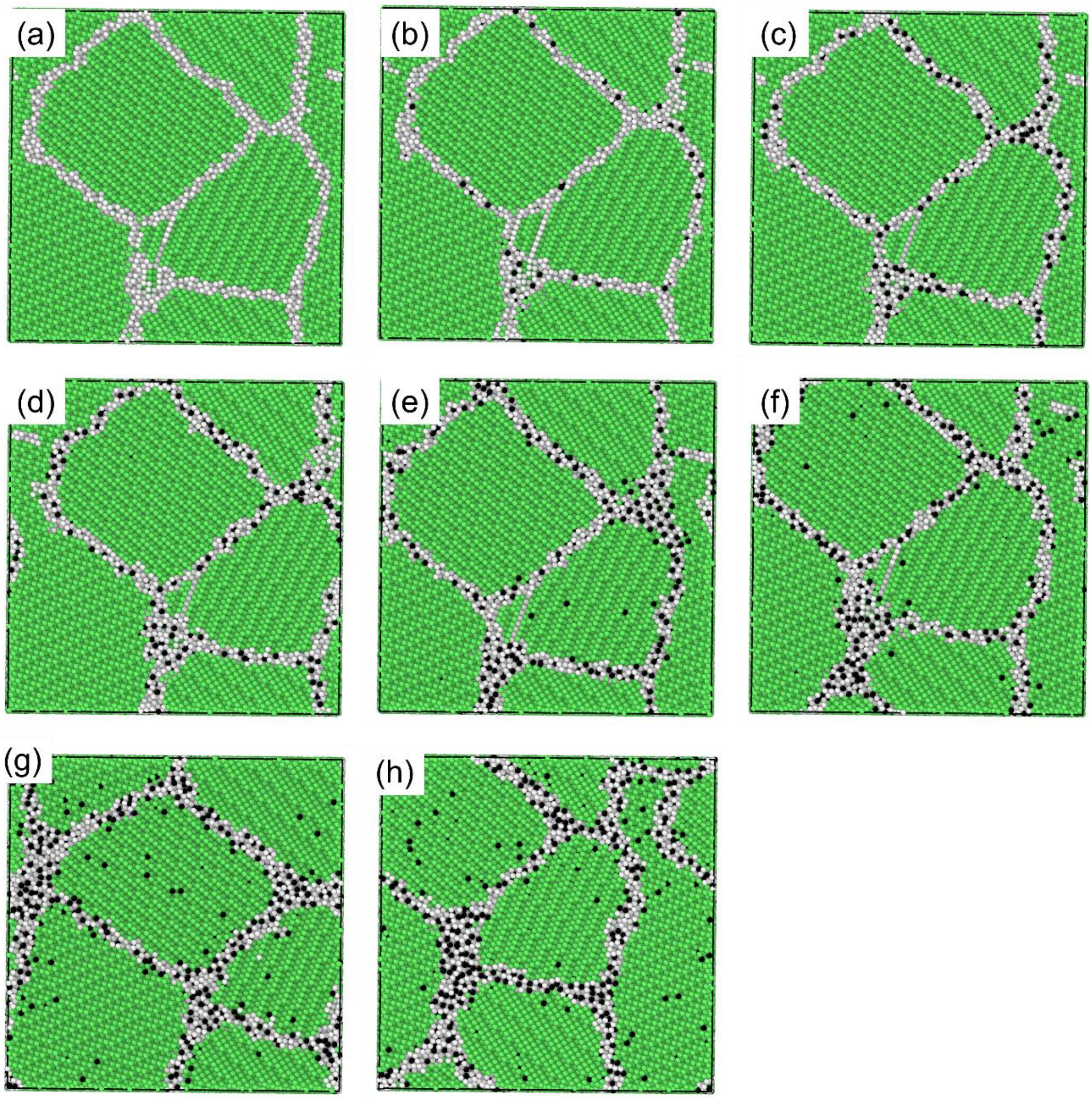

**Supplementary Fig. 1** (a) GB network without solute after full relaxation. (b)-(h) GB networks with the total solute concentrations ranging from 1 to 7 at.% after segregation at 300 K. GB evolution occurred after hybrid MD/MC simulations at each solute concentration level.

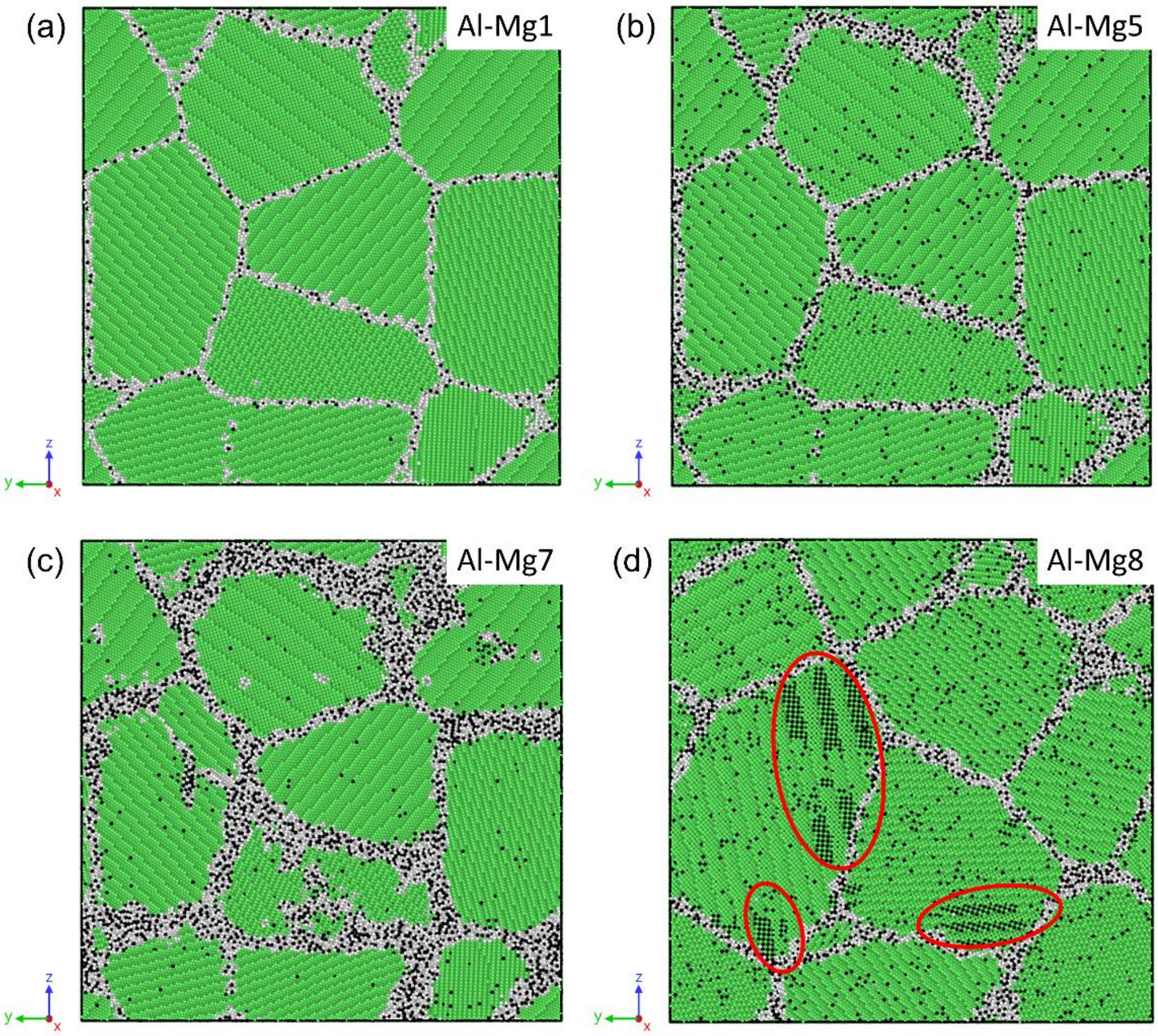

**Supplementary Fig. 2** The GB structures and solute distributions of the Al-Mg system with the dimension of $40 \times 40 \times 40\ nm^3$ at different Mg concentrations after hybrid MD/MC simulations at 300 K: (a) Al-Mg 1 at.%, (b) Al-Mg 5 at.%, (b) Al-Mg 7 at.%, and (b) Al-Mg 10 at.%. The green spheres are fcc atoms while the gray ones represent the GB atoms. The dark spheres are the solute atoms.

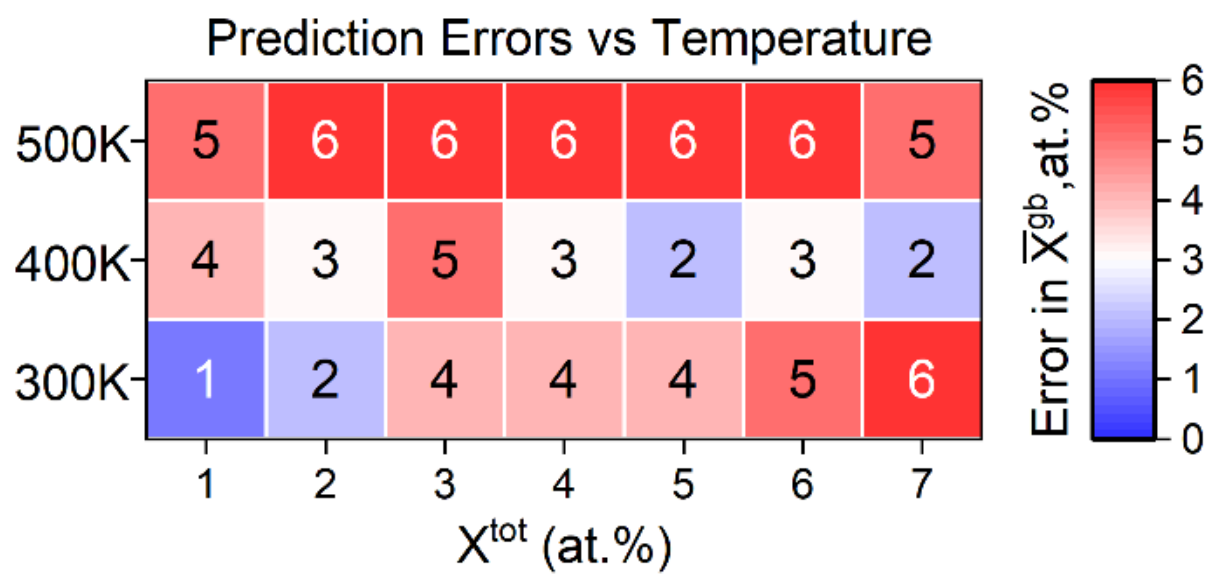

**Supplementary Fig. 3** Prediction errors of the DS model under different temperatures in the Al-16 system.

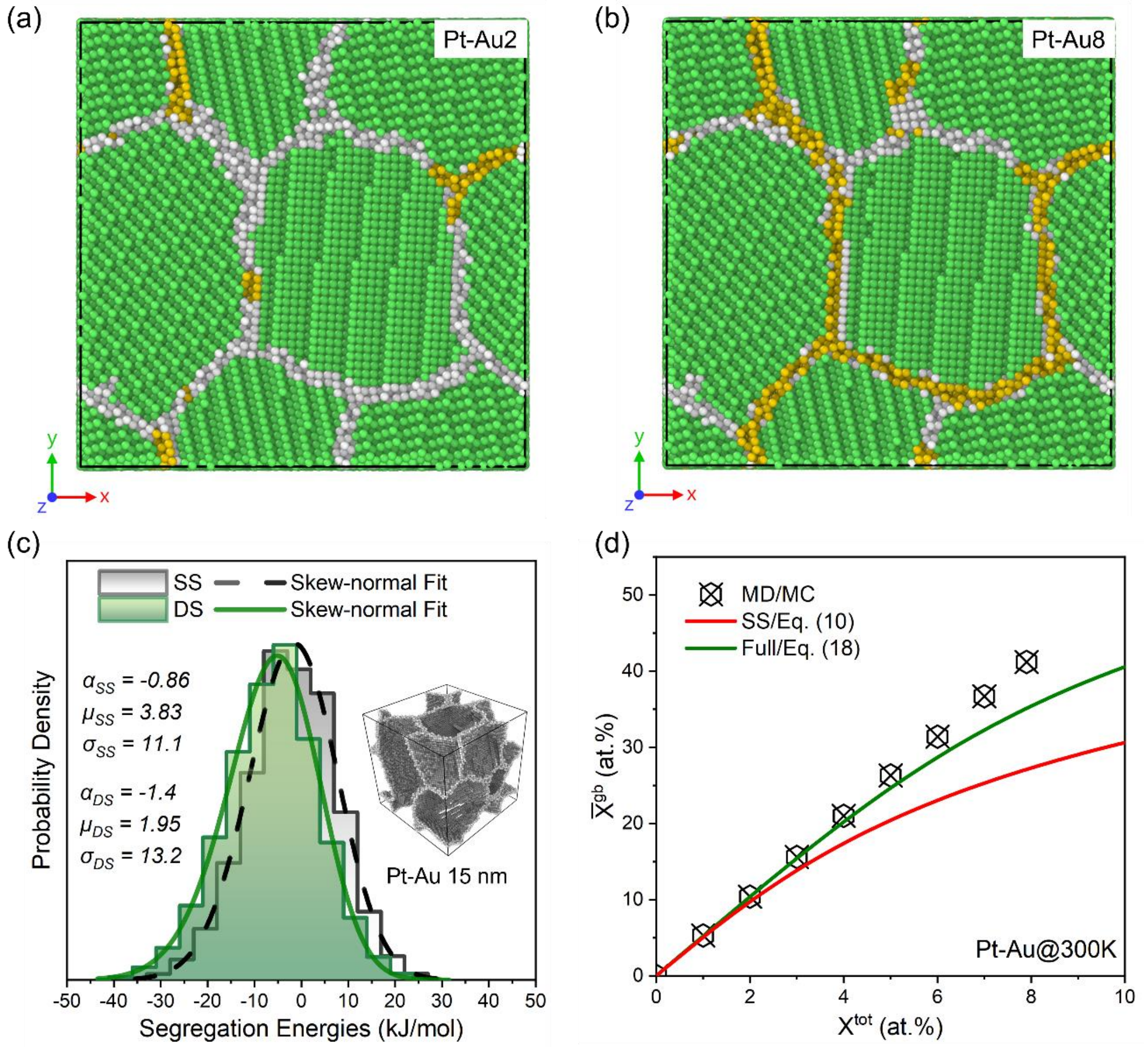

**Supplementary Fig. 4** The GB structures and solute distributions of the Pt-Au system after hybrid MD/MC simulations at 300 K: (a) Pt-Au 2 at.%, (b) Pt-Au 8 at.%. (c) The SS and DS segregation energy spectra of the Pt-Au system. The inset shows the thermally relaxed GB network of the pure NC Pt. (d) The SS and DS segregation prediction results against hybrid MD/MC results at 300 K.

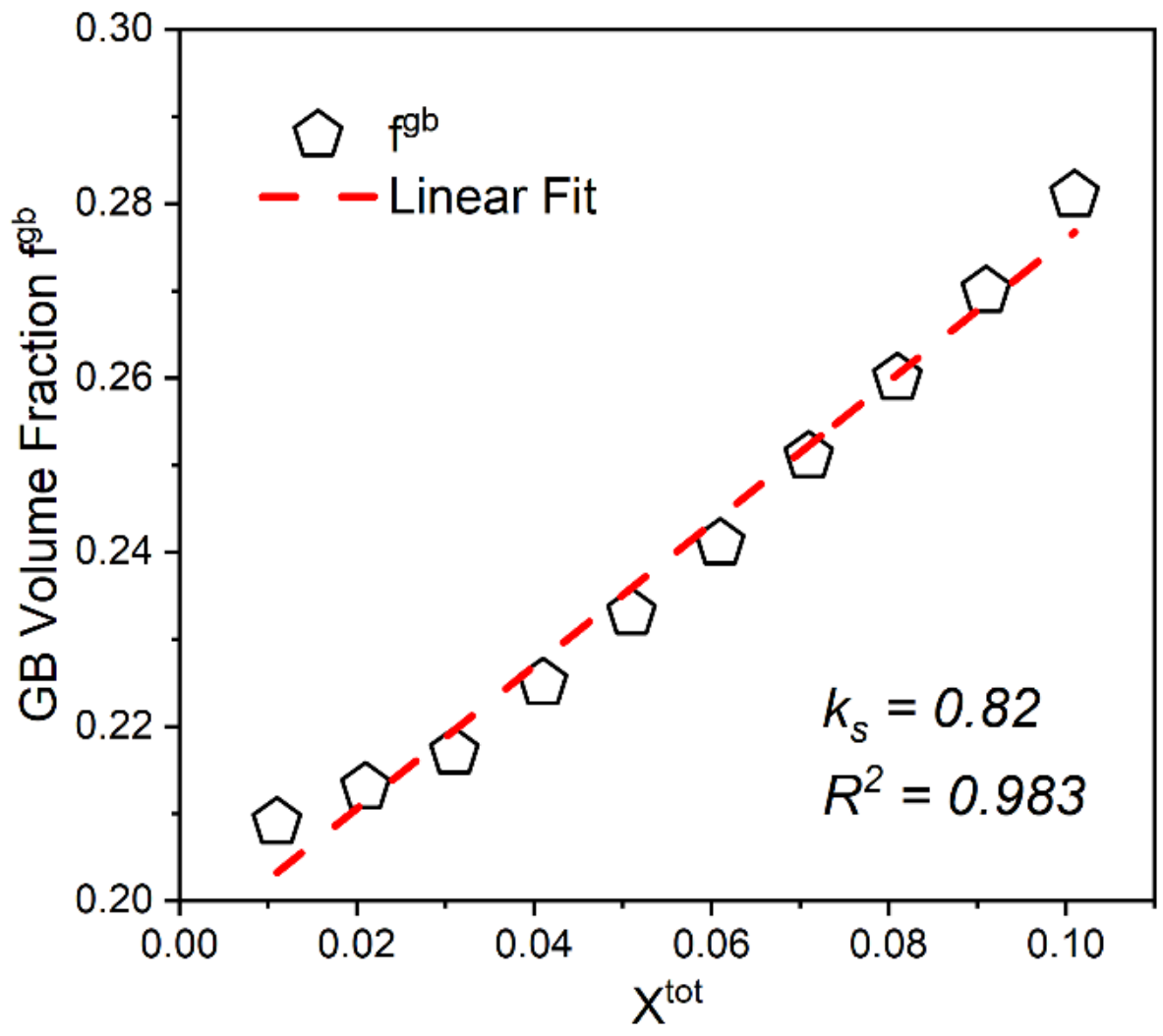

**Supplementary Fig. 5** Linear correlation between GB volume fraction $f^{gb}$ and $X^{tot}$ for the Ag-Ni system with the slop $k_s = 0.82$.

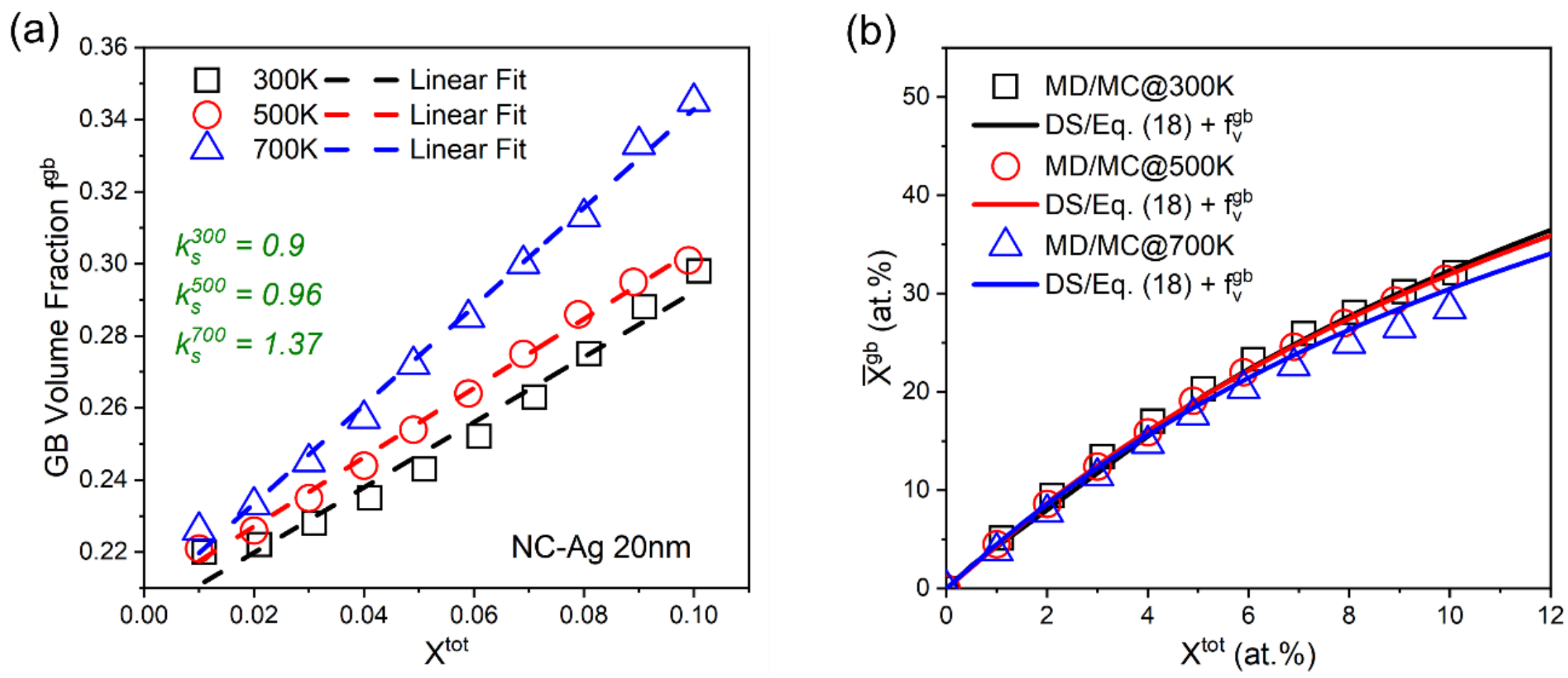

**Supplementary Fig. 6** (a) Correlation of the $X^{tot}$ and $f^{gb}$ at different temperatures. Obviously, the slope $k_s$ exhibits strong temperature dependence with higher slope at elevating temperatures. (b) The DS model prediction results, with $k_s = 1$ as a constant for the Ag-Ni system, align very well with the hybrid MD/MC results at different temperatures. This indicates that we can treat $k_s$ as a constant.

Supplementary Fig. 6(a) shows the correlation between $f^{gb}$ and $X^{tot}$ of the previously measured Ag-Ni system at 300, 500, and 700 K, with the size of $(20\ \text{nm})^3$. The NC-Ag 20nm model was taken from Ref. [1]. The results show that the slope $k_s$ increases with temperature form 0.9 at 300 K to 1.37 at 700 K. To simplify the slope $k_s$, we just approximate it with the value of $k_s = 1$. Then, we use the DS spectrum (from the model of Ag-16 nm) parameters in this work to predict the segregation states of the Ag-20 nm at different temperatures, as shown in Supplementary Fig. 6(b). The predicting results aligns perfect with the real segregation states at different temperatures. Therefore, we can just approximate the slope to 1 for different temperature conditions.